\documentclass[a4paper]{article}
\usepackage[text={6in,9in},centering]{geometry}
\usepackage{amsmath,amssymb}
\usepackage{multirow,multicol, makecell, booktabs}
\usepackage{hyperref}
\usepackage{cleveref}
\usepackage{lscape}
\usepackage{nccbbb}
\usepackage{caption}
\usepackage{subcaption}
\usepackage{booktabs}
\usepackage{longtable}
\usepackage{adjustbox}
\usepackage{authblk}

\newcolumntype{L}[1]{>{\raggedright\let\newline\\\arraybackslash\hspace{0pt}}m{#1}}
\newcolumntype{C}[1]{>{\centering\let\newline\\\arraybackslash\hspace{0pt}}m{#1}}
\newcolumntype{R}[1]{>{\raggedleft\let\newline\\\arraybackslash\hspace{0pt}}m{#1}}
\newcommand{\Qf}{\mathrm{Q_1}}
\newcommand{\Qt}{\mathrm{Q_3}}

\newcommand{\ApEn}{\mathrm{ApEn}}
\newcommand{\SampEn}{\mathrm{SampEn}}
\newcommand{\PE}{\mathrm{PE}}
\newcommand{\WPE}{\mathrm{WPE}}

\newcommand{\Dh}{\mathrm{D_H}}
\newcommand{\DB}{\mathrm{D_B}}
\newcommand{\LBP}{\mathrm{LBP}}
\newcommand{\LNDP}{\mathrm{LNDP}}
\newcommand{\LGP}{\mathrm{LGP}}
\newcommand{\FuzzEn}{\mathrm{FuzzEn}}
\newcommand{\DistEn}{\mathrm{DistEn}}
\newcommand{\DFA}{\mathrm{DFA}}
\newcommand{\IWMF}{\mathrm{IWMF}}
\newcommand{\IWBW}{\mathrm{IWBW}}

\newcommand{\ones}{\mathbf 1}
\newcommand{\diag}{\mathop{\bf diag}}
\newcommand{\ie}{{\it i.e.}}

\begin{document}

\title{A review of feature extraction and performance evaluation in epileptic seizure detection using EEG}

\author[1]{Poomipat Boonyakitanont}
\author[1]{Apiwat Lek-uthai}
\author[2]{Krisnachai Chomtho}
\author[1]{Jitkomut Songsiri}
\affil[1]{\small Department of Electrical Engineering, Faculty of Engineering, Chulalongkorn University, Thailand}
\affil[2]{\small Department of Pediatrics, Faculty of Medicine, Chulalongkorn University, Thailand}

\maketitle
\begin{abstract}
Since the manual detection of electrographic seizures in continuous electroencephalogram (EEG) monitoring is very time-consuming and requires a trained expert, attempts to develop automatic seizure detection are diverse and ongoing. Machine learning approaches are intensely being applied to this problem due to their ability to classify seizure conditions from a large amount of data, and provide pre-screened results for neurologists. Several features, data transformations, and classifiers have been explored to analyze and classify seizures via EEG signals. In the literature, some jointly-applied features used in the classification may have shared similar contributions, making them redundant in the learning process. Therefore, this paper aims to comprehensively summarize feature descriptions and their interpretations in characterizing epileptic seizures using EEG signals, as well as to review classification performance metrics. To provide meaningful information of feature selection, we conducted an experiment to examine the quality of each feature independently. The Bayesian error and non-parametric probability distribution estimation were employed to determine the significance of the individual features. Moreover, a redundancy analysis using a correlation-based feature selection was applied. The results showed that the following features --variance, energy, nonlinear energy, and Shannon entropy computed on a raw EEG signal, as well as variance, energy, kurtosis, and line length calculated on wavelet coefficients-- were able to significantly capture the seizures. An improvement of 4.77--13.51\% in the Bayesian error from the baseline was obtained.
\end{abstract}



\section{Introduction}
An epileptic seizure, as defined by the International League Against Epilepsy~\cite{fisher2014ilae}, is a temporary event of symptoms due to synchronization of abnormally excessive activities of neurons in the brain.
 It has been estimated that approximately 65 million people around the world are affected by epilepsy~\cite{thurman2011standards}.
Nevertheless, it is still a time-consuming process for neurologists to review continuous electroencephalograms (EEGs) to monitor epileptic patients.
Therefore, several researchers have developed different techniques that help neurologists to identify an epilepsy occurrence ~\cite{acharya2013automated,li2014using,kumar2015classification}. 
The whole process of automated epileptic seizure analysis primarily consists of data acquisition, signal pre-processing, feature extraction, feature or channel selection, and classification. This paper focuses on a selection of features commonly used in the literature, including statistical parameters (mean, variance, skewness, and kurtosis), amplitude-related parameters (energy, nonlinear energy, line length, maximum and minimum values) and entropy-related measures. These features can be categorized based on their interpretation or the domain from which the features are calculated. While some studies have considered a particular group of features applicable to their proposed classification method~\cite{gotman1982automatic,guo2010automatic,orosco2009epileptic}, others have applied various groups of features extracted from the time, frequency, and time-frequency domains. For example, 55 features were used with a support vector machine (SVM) and post-processing for neonatal seizure detection, which provided a good detection rate of 89.2\% with one false detection per hour~\cite{temko2011eeg}. Feature redundancy and relevance analysis were applied to 132 features to reduce the vector dimension~\cite{aarabi2006automated}. Fast correlation based-filter proposed in~\cite{yu2004efficient}, correlation-based selection (CFS)~\cite{hall1999correlation}, and ReliefF established in~\cite{kononenko1997overcoming}, were utilized to select non-redundant features and the filtered features were classified by an artificial neural network (ANN). It turned out that 30-optimal selected features by ReliefF achieved the best result with 91\% sensitivity and 95\% specificity. These studies showed that there is the need to review feature selection as relevant features are directly related to the seizure classification performance.

There have been a number of review articles focusing on automated epileptic seizure detection (AESD). A brief investigation on the AESD methods with some mathematical descriptions was provided in~\cite{acharya2013automated}.
The EEG analyses were categorized into time domain, frequency domain, time-frequency domain, and nonlinear methods.
The authors then discussed the importance of analysis of surrogate data and methods of epilepsy classifications of two (normal and ictal) and three (normal, interictal, and ictal) classes, where the nonlinear features outperformed the others with an accuracy of more than 99\%. However, implementation of statistical parameters was not addressed in this paper.
Applications of using entropy for epilepsy analysis were reported in~\cite{acharya2015application}.
Many types of entropy have been utilized in the AESD, both jointly and independently.
All entropies were mathematically stated in combination with their advantages and disadvantages. 
The authors also showed differences of each feature in the normal and ictal groups via the \textit{F}-value.
The three highest ranked features were Renyi's, sample, and spectral entropies.
Moreover, the seizure detection techniques of the whole process, including feature extraction, feature selection, and classification methods, were summarized~\cite{ahmad2016seizure}.
Feature extraction methods were divided into two spectral domains, \ie, time-frequency analysis and temporal domain. However, only a brief demonstration of each method was depicted; there was no explanation in the mathematical formulae and the AESD implementation was not described in the paper.
Characteristics of focal and non-focal seizure activities observed from EEG signals were also reviewed in~\cite{acharya2018characterization}.
Focal EEG signals and non-focal EEG signals were obtained from the epilepsy- affected and unaffected areas, respectively.
Nonlinear features, such as Hjorth parameters, entropy, and fractal dimension, were used to characterize and compare the focal and non-focal EEG signals using the Bern-Barcelona EEG database~\cite{andrzejak2012nonrandomness}.
Nevertheless, none of these features were expressed mathematically.

In addition, some studies have, in part, involved the AESD application.
Studies on channel selection techniques for processing EEG signals and reducing feature dimensions in seizure detection and prediction are needed because considering every channel may cause overfitting problems~\cite{alotaiby2015review}.
The channel selection techniques of filtering, wrapper, embedded, hybrid, and human-based techniques were graphically explained via flowcharts and found to show a high accuracy and low computational cost function.
Uses of deep learning, including ANNs, recurrent neural networks (RNNs), and convolutional neural networks (CNNs), in biological signal processing, including EEG, electromyography, and electrocardiography signals, were also reviewed in~\cite{faust2018deep}. The author summarized that there were 24 studies relevant to EEG signal analysis; two of which were focused on AESD and its prediction.
It was also reported that applying CNN on the AESD had an 86.9\% accuracy.
Furthermore, methods for non-stationary signal classification using time-frequency and time-scale representations, and their differences were discussed in depth in the seizure detection application~\cite{boashash2018designing}.
Several features were extracted after time-frequency and time-scale transformations: spectrogram, Extended Modified B-Distribution, compact kernel distribution (CKD), Wigner-Ville distribution (WVD), scalogram, Affine WVD (AWVD), pseudo-smoothed AWVD, and CKD with feature selection, fed to a random forest (RF).
The results showed that every time-frequency and time-scale transformation achieved an accuracy of more than 80\%, while CKD with 50 optimal selected features yielded an accuracy of 86.41\%.

A review mainly focusing on using the wavelet-based method for computer-aided seizure detection was described in~\cite{faust2015wavelet}.
The authors summarized approaches for seizure detection using different wavelets, \ie, discrete wavelet transform (DWT) and continuous wavelet transform (CWT), with the use of nonlinear-based features and chaos-based measurements.
The authors showed that the Daubechies wavelet tap 4 was the most favorable wavelet for use in seizure detection.
However, there was no explanation about the benefits of each wavelet, no mathematical expression of wavelet transforms, and no comparison of the results between using different wavelet-based techniques.
In addition, a review of epileptic seizure prediction processes, including data collection, signal pre-processing, feature extraction, feature selection, classification, and performance evaluation, was reported in~\cite{assi2017towards}. In the feature extraction, the authors categorized features into four groups of linear univariate, linear bivariate, nonlinear univariate, and nonlinear bivariate features. In this case, a nonlinear measure was defined as an attribute related to dynamic states of the brain. Moreover, univariate and bivariate features were extracted from individual and multiple channels, respectively. However, no mathematical description of all the processes was given.

From the past literature, we conclude that selecting distinguished features is required for applying machine learning (ML) approaches to AESD, but previous studies still lack some conclusive points. Firstly, many previous papers have applied a combination of features while providing neither a clear conclusion about which feature has the highest contribution nor an intuitive meaning of using such features~\cite{temko2011eeg,alickovic2018performance,ammar2016seizure}. Secondly, mathematical expressions of some complicated features, such as entropies, and Hurst exponent (HE), were not consistently stated and so may lead to an incorrect implementation~\cite{acharya2013automated,ahmad2016seizure}. Thirdly, most papers focused on a particular group of features that use the classification accuracy as the main indicator to conclude the effectiveness of those features~\cite{ammar2016seizure,anand2017automatic,chakraborti2018machine}. However, we believe that this cannot be a fair indicator for a method comparison. To illustrate this, we consider using a default classifier that regards all signals as normal. This method already yields a high accuracy if the data set in consideration contains mostly normal signals, known as imbalanced data sets. Hence, in our opinion, a performance comparison of all features is required using the same experimental setting based on the Bayes classifier and should be done with a well-adjusted classification indicator. This paper, therefore, aims to explore all those issues by providing a review of the feature extraction process and investigating the contribution of each feature used in the seizure classification in EEG signals. We reported the accuracies and other related metrics (if available) of previous studies on the same EEG databases. Our conclusions related to feature significance are drawn from the experiments using commonly-used features and on a public database reported in the literature.

\paragraph{Literature search} In this review, we mainly gathered articles via Google Scholar to obtain publications from several databases, including ScienceDirect and IEEE Xplore.
Studies that were published between January 2010 and December 2018 were primarily considered to reveal the current status of work.
The first search aimed to find review articles using the following keywords: review, automated epileptic seizure detection, EEG signal, channel selection, feature selection, deep learning, and feature extraction.
In the second search, we primarily used the keywords: feature, nonlinear feature, feature extraction, EEG signal, seizure detection, epilepsy detection, automated epileptic seizure detection, entropy, neural network, and fractal dimension for finding research articles focusing on the feature extraction.
In addition, we also included more articles from ScienceDirect and Mendeley recommender systems and references of the included studies.
Studies that focused on animals, devices, software, medical treatment, drugs, and surgery were excluded.
Only relevant studies that used only EEG signals, applied non-typical features or combinations of widely used features, and had a well-described feature explanation were included.
This led to a total of 55 papers, including nine review studies, to be thoroughly reviewed on the feature extraction. The other references were related to techniques used in feature extraction.

This manuscript is organized as follows.
\Cref{sec:feature_extraction} describes the features extracted from each domain and their computational complexity.
Features with detailed mathematical description are categorized according to the domains from which they were calculated: time, frequency, and time-frequency domains. In \Cref{sec:use_feature}, we discuss evaluation metrics and summarize methods for the AESD based on the use of features. \Cref{sec:method_eva} explains the Bayes method for classification and CFS for assessing feature performance and redundancy, and the results are compared in \Cref{sec:exp}.

\section{Feature extraction}
\label{sec:feature_extraction}
This section describes the details of features commonly used in the literature of EEG seizure detection, categorized by feature domains: time, frequency, and time-frequency domains. Time-domain features (TDFs) are those calculated on raw EEG signals or on pre-processed signals done in the time domain, such as empirical mode decomposition (EMD). Frequency-domain features (FDFs) are computed on discrete-Fourier transform of raw EEG signals. Time-frequency-domain features (TFDFs) are defined on transformed EEG signals that contain both time and frequency characteristics, such as short-time Fourier transform (STFT) spectrogram or DWT. In the last section, we briefly explain the computational complexity of each feature so that users can take this concern when performing a real-time implementation of AESD. To what follows, we denote $X = \left[x[0], x[1],\ldots,x[N-1] \right]$ a sequence of length $N$ used for extracting a feature. For instance, $X$ can be an epoch of a raw EEG segment, absolute values of a raw EEG segment, power spectral density (PSD), approximation or detail coefficients from any wavelet transform, or intrinsic mode functions (IMFs) from EMD.
\subsection{Time-domain features (TDFs)}
\label{sec:feature_time}
A TDF is calculated from a raw EEG signal or a decomposed signal performed on a time domain. In the literature, one example of signal transformation is EMD that decomposes a signal into IMFs \cite{huang2014hilbert} and is commonly applied to signals that apparently exhibit non-stationary properties. In this section, well-known TDFs are briefly described and those involving an entropy concept are explained with mathematical expressions. 
\begin{enumerate}
	\item Groups of statistical parameters have been frequently used to discriminate between ictal and normal patterns, because it is assumed that EEG statistical distributions during a seizure and normal periods are different. These parameters are \emph{mean, variance, mode, median}, \emph{skewness} (third moment describing data asymmetry), and \emph{kurtosis} (fourth moment determining tailedness of the distribution). The \emph{minimum} and \emph{maximum} values are also used to quantify the range of data or the magnitude of signal baseline. Other statistical parameters include \emph{coefficient of variation (CV)} defined as the ratio of the standard deviation (SD) to the sample mean that explains the dispersion of the data in relation to the population mean; \emph{first $(\Qf)$ and third quartiles $(\Qt)$} that quantify the data denseness; and  \emph{interquartile range (IQR)} that measures a deviation between the first and third quartiles.
	\item \emph{Energy}, \emph{average power}, and \emph{root mean squared value (RMS)} are mutually relevant to amplitude measurements. The energy is a summation of a squared signal, the average power is the signal mean square, and the RMS is the square root of the average power.
	\item \emph{Line length}, sometimes called curve length, is the total vertical length of the signal defined as
\begin{equation}
\mathrm{L}(X) = \sum\limits_{i=1}^{N-1} \lvert x[i]-x[i-1] \rvert.
\label{eq:line_length}
\end{equation}
It was originally presented in \cite{esteller2001line} as an approximation of Katz's fractal dimension.
	\item \emph{Amplitude-integrated EEG} (\emph{aEEG}) is a visual inspection that is widely used for seizure detection in hospitals \cite{hellstrom2008atlas}. The aEEG signal is obtained by calculating the difference between adjacent maximum and minimum values, \ie, the peak-to-peak rectification. Consequently, the rectified signal is passed through the process of interpolation and semi-logarithmic compression proposed in \cite{maynard1969device}. The interpolated signal is linearly displayed in the range 0--10 $\mathrm{\mu}$V and semi-logarithmically compressed in the range over 10 $\mathrm{\mu}$V. This feature is applied based upon the assumption that during seizure events the amplitude of signals are shifted up from its baseline during normal events. 
	\item \emph{Nonlinear energy} (\emph{NE}), firstly established in~\cite{kaiser1990simple}, extends the concept of energy (quadratic measure) to including indefinite terms of shifted and lagged sequences, defined as 
\begin{equation}
\mathrm{NE}(X) = \sum\limits_{i=1}^{N-2}\left(x^2[i]-x[i+1]x[i-1]\right).
\label{eq:nonlinear_energy}
\end{equation}
In~\cite{kaiser1990simple}, if a signal has a simple harmonic motion with the amplitude $A$ and the oscillation frequency $\omega$, it can be derived that the NE is proportional to $A^2\omega^2$ when the sampling frequency is high. Hence, high values of NE can indicate both shifted values in a high frequency of oscillation and amplitude.
	\item \emph{Shannon entropy} (\emph{ShEn})~\cite{shannon1948mathematical} reflects the uncertainty in random process or quantities. It is defined as	
\begin{equation}
\mathrm{ShEn}(X) = -\sum\limits_i p_i \log p_i,
\label{eq:ShEn}
\end{equation}
where $p_i$ is the probability of an occurrence of each of value in $X$.
	\item \emph{Approximate entropy} (\emph{ApEn})~\cite{pincus1991approximate} is a measure of the regularity and fluctuation in a time series derived by comparing the similarity patterns of template vectors.
	The \emph{template vector} of size $m$ is defined as a windowed signal: $u[i] = 
\begin{bmatrix}
	 x[i] & x[i+1] & \cdots & x[i+m-1]
\end{bmatrix}^T
$,
and we first consider the self-similarity of the template vector $u[i]$ with a tolerance $r$, defined by
\begin{equation*}
C_i^m(r) = \frac{1}{N-m+1}\sum\limits_{j=0}^{N-m}\Theta\left(r-\lVert u[i]-u[j] \rVert_{\infty}\right),
\label{eq:ApEn_match}
\end{equation*}
where $\Theta\left(x\right)$ is the Heaviside step function, \ie, $\Theta\left(x\right)$ is one when $x\geq 0$, and zero otherwise. When $X$ is mostly self-similar, then $u[i]$ and $u[j]$ sequences are very close and thus $C_i$ is high. 
The ApEn aggregates the self-similarity indices over all shifted possibilities of template vectors given template length and tolerance. The ApEn is defined as 
\begin{equation}
\ApEn\left(X,m,r\right) = \frac{1}{N-m+1}\sum\limits_{i=0}^{N-m} \log C_i^m(r)-\frac{1}{N-m}\sum\limits_{i=0}^{N-m-1} \log C_i^{m+1}(r).
\label{eq:ApEn}
\end{equation}
	\item \emph{Sample entropy} (\emph{SampEn})~\cite{richman2000physiological} is based upon a concept similar to the ApEn, where the SampEn compares the total number of template vectors of size $m$ and $m+1$. The SampEn differs from the ApEn in that the self-similarity of \emph{all pairs} of template vectors $u[i]$ and $u[j]$ with a tolerance $r$ is calculated by
\begin{equation*}
\phi^m(r) = \sum\limits_{j=0,j\neq i}^{N-m}\sum\limits_{i=0}^{N-m}\Theta\left(r-\lVert u[i]-u[j] \rVert_{\infty}\right).
\end{equation*}
If the signals are self-similar, $\phi^m(r)$ is high. The SampEn is defined by
\begin{equation}
\SampEn\left(X,m,r\right) =\log \phi^m(r) - \log \phi^{m+1}(r).
\label{eq:SampEn}
\end{equation}
	\item \emph{Permutation entropy} (\emph{PE})~\cite{bandt2002permutation} is a measure of the local complexity in a signal. With the template vector $u[i]$ and a permutation pattern $\pi_k$ of order $m$ consisting of $m!$ patterns, the permutation pattern probability for all $k=1,2,\ldots,m!$ is defined as the probability that a template vector has the same pattern as the permutation pattern:
\begin{equation}
p(\pi_k) = \frac{1}{N-m+1}\sum\limits_{i=0}^{N-m} f\left(u[i],\pi_k\right),
\label{eq:PE_match}
\end{equation}
where $f\left(u[i],\pi_k\right) = 1$ when $u[i]$ and $\pi_k$ are the same pattern, and zero otherwise.
In this case, the pattern is defined by the order of $u[i]$ corresponding to its element.
Then, the PE is  defined as
\begin{equation}
\PE\left(X\right) = -\sum\limits_{k=1}^{m!}p(\pi_k)\log p(\pi_k).
\label{eq:PE}
\end{equation}
	\item \emph{Weighted-permutation entropy} (\emph{WPE})~\cite{fadlallah2013weighted} is a measure of the process complexity extended from PE by putting the variance of the template vector as a weight ($w_i$) in computing the probability of each permutation pattern, as calculated by
\[
p(\pi_k) = \left. \left ( \sum\limits_{i=0}^{N-m} f\left(u[i],\pi_k\right)w_i \right )  \middle / \left ( \sum\limits_{i=0}^{N-m}w_i \right ) \right.,
\]
where $f\left(u[i],\pi_k\right) = 1$ when $u[i]$ and $\pi_k$ are the same pattern, and zero otherwise.
Finally, the WPE is calculated by
\begin{equation}
\WPE\left(X\right) = -\sum\limits_{k=1}^{m!} p(\pi_k)\log p(\pi_k).
\label{eq:WPE}
\end{equation}
\item \emph{Fuzzy entropy} (\emph{FuzzEn}) is another measure of the process uncertainty, where its calculation is based on a zero-mean sequence~\cite{chen2007characterization}. Consider a windowed signal with mean removed: $ u[i] = 
\begin{bmatrix}
	 x[i] & x[i+1] & \cdots & x[i+m-1]
\end{bmatrix}^T - \bar{x}[i]\ones,
$
where $\bar{x}[i] = (1/m)\sum\limits_{j=0}^{m-1}x[i+j]$, and $\ones$ is a column vector of ones. Like other entropy measures, the sequence $u$ is used to compute a self-similarity index, $\phi$, but for FuzzEn, a Gaussian function is used as defined by
\begin{equation*}
\phi^{(m)}(r) = \frac{1}{(N-m)(N-m-1)}\sum\limits_{j=0,j\neq i}^{N-m}\sum\limits_{i=0}^{N-m}e^{d_{ij}^2/2r^2 } ,
\label{eq:FuzzyEn_sim}
\end{equation*}
where $d_{ij} = \lVert u[i]-u[j] \rVert_{\infty}$. If $X$ is self-similar, then $d_{ij}$ is small and relative to the Gaussian curve with variance $r^2$, and so we can determine how high $\phi^{(m)}(r)$ is. The FuzzEn is then defined to reflect the change of $\phi$ as the window length $m$ is increased to $m+1$,
\begin{equation}
\FuzzEn\left(X,m,r\right) = \log \phi^{m}(r)-\log \phi^{m+1}(r).
\label{eq:FuzzyEn}
\end{equation}
	\item \emph{Distribution entropy} (\emph{DistEn}) measures the uncertainty of a process by estimating the probability distribution with a histogram~\cite{li2015assessing}.
With the zero-mean vector defined previously in FuzzEn, the distance between two template vectors (windowed signals) is $d_{ij} = \lVert u[i]-u[j] \rVert_{\infty}$ and is used to construct a histogram with $B$ bins. The DistEn is then defined to be the entropy of the random process $d_{ij}$:
\begin{equation}
\DistEn\left(X,m,B\right) = -\frac{1}{\log B}\sum\limits_{i=1}^B p_i\log p_i,
\label{eq:DistEn}
\end{equation}
where $p_i$ is the relative frequency of each bin. A high DistEn value reflects the unpredictable nature of the original sequence $X$. 
	\item \emph{Singular Value Decomposition (SVD) entropy (SVDEn)}, first established in a brain-computer interface application, was proposed to measure the temporal and spatial complexity of a process based on entropy \cite{roberts1999temporal}. Any rectangular matrix $A$ has an SVD of $A = U\Sigma V^T$, where $\Sigma = \diag(\sigma_1,\sigma_2,\ldots,\sigma_m)$ containing the singular values of $A$. The number of non-zero singular values of $A$ determines its rank. In feature extraction, the normalized singular value is given by $\tilde{\sigma}_j = \sigma_j/\sum\limits_i^m \sigma_i$ and the SVDEn is defined as the entropy form of these normalized singular values:
\begin{equation}
	\mathrm{SVDEn} (X) = -\sum\limits_{j} \tilde{\sigma}_j\log \tilde{\sigma}_j.
\label{eq:svd_entropy}
\end{equation}
For the temporal measurement, a delay method of time $\tau$ is used to construct each row of the matrix $A$. For the spatial SVDEn, each row of the matrix $A$ is a sequence of the EEG signal from each channel. It is claimed that when the EEG signals are noisy, potentially all singular values are non-zero ($A$ is full rank). More importantly, it is observed that singular values associated with noisy signals are significantly smaller than the deterministic (noiseless) signal. Therefore, a low SVDEn value spatially and temporally indicates patterns of epileptic seizure in EEG signals because the singular values from the EEG signals from a seizure are higher than those of the normal signals.

	\item \emph{Hurst exponent} (\emph{HE})~\cite{hurst1951long} indicates a degree of time series tendency. We call a \emph{partial time series} of $X$ as a chopped signal having length $m$, such that $m < N$, \emph{e.g.} $m=N/2$. A \emph{cumulative deviate series}, or so-called \emph{partial cumulative time series}, is defined as 
\begin{equation}
z[t] = \sum\limits_{j=0}^{t} \left(x[j]-\bar{x}\right),
\label{eq:cumulat_series}
\end{equation}
where $\bar{x} = (1/m) \sum\limits_{i=0}^{m-1}x[i]$ is the average of the partial time series. With these auxiliary signals, we consider $R(m)$ as a cumulative range: $R(m) = \max\limits_{t} z[t] - \min\limits_{t} z[t]$, and $S(m)$ as the SD of the partial time series. Then, the HE is defined as the slope of the straight line that fits $\log \left(R(m)/S(m)\right)$ as a function of $\log m$ by the least-square method.
\[
\log \left(R(m)/S(m)\right) = \mathrm{HE} \cdot \log m. 
\]
With a fixed $m$, we observe how the signal is fluctuated from its mean via $z$ and the range of this fluctuation is explained in $R$, relative to the SD (normalized by $S$). Thus, the HE tends to capture how this range grows as $m$ increases and approximates it as a linear growth in a log-scale.
	\item \emph{Fractal dimension} is a mathematical index for measuring signal complexity \cite{falconer2004fractal}.
	The Higuchi dimension ($\Dh$) was proposed to calculate the fractal dimension for a time series in the time domain \cite{higuchi1988approach}.
	A partial length, denoted as $L_m(\tau)$, is calculated from a partial time series with a time interval $\tau$ and an initial time $m=0,1,2,\ldots,\tau$ as
\begin{equation}
	L_m(\tau) = \frac{1}{\tau}\left(\sum\limits_{i=1}^{\lfloor\frac{N-m}{\tau}\rfloor}\lvert x[m+i\tau]-x[m+(i-1)\tau] \rvert\right)\frac{N-1}{\lfloor\frac{N-m}{\tau}\rfloor \tau}.
\label{eq:fractal_length}
\end{equation}
The slope of the straight-line fitting $\log L(\tau)$ as a function of $\log\tau$ is equal to $-\Dh(X)$, where $L(\tau)$ is the average of $L_m(\tau)$ over $\tau$.
Moreover, there is another approach, the Minkowski dimension or box-counting dimension ($\DB$), that measures the complexity of images and signals by counting square boxes that cover the object \cite{falconer2004fractal}.
Here, $\mathrm{D_B}(X)$ is defined as
\begin{equation}
	\DB(X) = \lim\limits_{\epsilon \to 0} \frac{\log N(\epsilon)}{\log(1/\epsilon)},
\label{eq:box_counting}
\end{equation}
where $N(\epsilon)$ is the number of boxes of size $\epsilon$ of each side required to cover the signal.
	\item \emph{Local Binary Pattern} (\emph{LBP}), \emph{Local Neighbor Descriptive Pattern} (\emph{LNDP}), and \emph{Local Gradient Pattern} (\emph{LGP}) are encoding techniques that transform a partial time series into encoded patterns \cite{kumar2015classification,jaiswal2017local}.
For $i=0,1,\ldots,N-m-1$ when $m$ is a positive even integer, the LBP of the partial time series is defined as the difference between the partial time series and its center:
\begin{equation}
\LBP_i(X) = \sum\limits_{j=0}^{m/2-1} \Theta\left(x[i+j]-x[i+m/2]\right)2^j+\sum\limits_{j=m/2}^{m-1} \Theta\left(x[i+j+1]-x[i+m/2]\right)2^j,
\label{eq:1D-LBP}
\end{equation}
where $\Theta(x)$ is the Heaviside step function.
	The LNDP captures the relationship of signals from the neighbor sequences and is defined by
\begin{equation}
\LNDP_i(X) = \sum\limits_{j=0}^{m-1} \Theta(x[i+j]-x[i+j+1])2^{j}.
\label{eq:LNDP}
\end{equation}
Whereas the LGP represents the changes in a signal as measured by the difference between signal gradients and the average gradient, and defined by 
\begin{multline}
\LGP_i(X) = \sum\limits_{j=0}^{m/2-1} \Theta\left(\lvert x[i+j]-x[i+m/2]\rvert-\frac{1}{m}\sum\limits_{k=0}^{m}\lvert x[i+k]-x[i]\rvert\right)2^{j}\\
+\sum\limits_{j=m/2}^{m-1} \Theta\left(\lvert x[i+j+1]-x[i+m/2]\rvert-\frac{1}{m}\sum\limits_{k=0}^{m}\lvert x[i+k]-x[i]\rvert\right)2^{j}.
\label{eq:LGP}
\end{multline}
These three parameters are pattern indicators and their histograms can form into features. The main advantage of these parameters is their invariant properties. The LBP is globally invariant, while the LNDP and LGP are not affected by local variation in the signal. From the algorithms, it was observed that histograms from normal and seizure EEG signals are different; the histograms of LBP, LNDP, and LGP from the abnormal signals have a high frequency at some specific values.
	\item Three \emph{Hjorth parameters} (\emph{activity, mobility}, and \emph{complexity}) were established to characterize the spectral properties of EEG signals in the time domain~\cite{hjorth1970eeg}. The activity is the variance of a signal and is high during seizure events in terms of a high amplitude variation from its mean. The mobility is a measure of a quantity related to the SD of the signal PSD. The mobility is expressed as the ratio of the SD of a signal derivative to the SD of the signal. Complexity represents the difference between the signal and a pure sine wave. The complexity is defined by the ratio of the mobility of a signal derivative to the mobility of the signal.
	
	\item \emph{Detrended fluctuation analysis} (\emph{DFA}) is a method of examining the self-correlation of a signal, which is similar to the HE. Its calculation is conducted by the following steps \cite{bryce2012revisiting}.
	A \emph{partial cumulative time series}, $z[t]$, as defined in \eqref{eq:cumulat_series} is used to determine a trend of the signal.
	Subsequently, $z[t]$ is then segmented into a signal of length $n$.
	A local trend of the segment is estimated by a least-square straight line $\hat{z}[t]$.
	The mean-squared residual $F[n]$ is then obtained from
\begin{equation}
	F[n] = \sqrt{\frac{1}{N}\sum\limits_{t=0}^{N-1}\left(z[t]-\hat{z}[t]\right)^2}.
\label{eq:DFA}
\end{equation}
$\DFA(X)$ is defined as the slope of the straight-line fitting $\log F[n]$ as a function of $\log n$ in a regression setting. The result indicates the fluctuation of the signal when its local trend is removed.
	\item \emph{Number of zero-crossings} is an indirect measurement of the frequency characteristics of a signal. If this number is large, it means the signal contains high frequency components and more uncertainty.

	\item \emph{Number of local extrema} is the total number of local maxima and minima in a signal.
	It is similar to the number of zero-crossings that indirectly represents the frequency measurement of the signal.
\end{enumerate}

\subsection{Frequency-domain features (FDFs)}
\label{sec:feature_freq}
Frequency domain analysis is also crucial, since a frequency representation of an EEG signal provides some useful information about patterns in the signal.
The PSD and the normalized PSD (by the total power) are mostly used to extract features that represent the power partition at each frequency. This section describes features that are extracted from the PSD.
\begin{enumerate}
	\item The \emph{energy} is extracted from some specified frequency ranges, usually corresponding to normal EEG activities, to determine the EEG rhythmicity in each frequency range.
	\item \emph{Intensity weighted mean frequency} (\emph{IWMF}), also known as the mean frequency, gives the mean of the frequency distribution using the normalized PSD, and is defined as
\begin{equation}
\IWMF\left(X\right) = \sum\limits_{k}x[k] f[k],
\label{eq:iwmf}
\end{equation}
where $x[k]$ is the normalized PSD of an EEG epoch at the frequency $f[k]$.
	\item \emph{Intensity weighted bandwidth} (\emph{IWBW}), or the SD frequency, is a measure of the PSD width in terms of the SD, and is defined as 
\begin{equation}
\IWBW\left(X\right) = \sqrt{ \sum\limits_{k}x\left[k\right]\left(f[k]-\mathrm{IMWF}\left(X\right)\right)^2},
\label{eq:iwbw}
\end{equation}
where $x[k]$ is the normalized PSD. According to the seizure patterns in typical EEG signals, the PSD is sharper during seizure activities. Therefore, the IWBW is smaller during those activities.
	\item \emph{Spectral edge frequency} (\emph{SEF}), denoted as $\mathrm{SEF}\alpha$, is the frequency at which the total power of the normalized PSD, $x[k]$, is equal to $\alpha$ percent. We define $\mathrm{SEF}\alpha$ as the frequency $f[k_{\mathrm{SEF}}]$ such that $\sum\limits_{k=0}^{k_{\mathrm{SEF}}} x[k] = 0.01\alpha$. By this definition, the \emph{median frequency} is a related feature, defined as the frequency at which the total power is equally separated and is, therefore, equal to the SEF50.
	\item \emph{Spectral entropy} (SE) is a measure of the random process uncertainty from the frequency distribution.
A low SE value means the frequency distribution is intense in some frequency bands.
Its calculation is similar to that for the ShEn but replaces the probability distribution with the normalized PSD as follows:
\begin{equation}
\mathrm{SE}(X) = -\sum\limits_k x[k] \log x[k].
\label{eq:SE}
\end{equation}
	\item \emph{Peak frequency}, also called dominant frequency, is the frequency at which the PSD of the highest average power in its full-width-half-max (FWHM) band has the highest magnitude. Since the peak frequencies of normal and seizure are located differently, it can be used to differentiate the two events.

	\item \emph{Bandwidth} of the dominant frequency is defined as the FWHM band corresponding to the peak frequency.
Regarding the rhythmicity of an EEG signal, it can be implied that, during seizure activities, there are frequency ranges that are more intense. Hence, the bandwidth is supposedly small during the seizure activities.
	\item \emph{Power ratio} of the current and background epochs at the same frequency range is determined to compare the powers. The power of a seizure period is relatively higher than that of the background.
\end{enumerate}

\subsection{Time-frequency-domain features (TFDFs)}
\label{sec:feature_time_freq}
Since an seizure occurrence can be captured by both TDFs and FDFs, many transformation and decomposition techniques that give information in terms of time and frequency have been widely considered, \emph{e.g.}, STFT spectrogram and DWT analyses
In this section, all the features are computed from the results of decomposition techniques. For example, when the DWT was used, features were calculated from the decomposition coefficients of some specific levels.
\begin{enumerate}
	\item Statistical parameters are also common in the time-frequency domain. Like TDFs, these attributes are computed to distinguish statistical characteristics of normal and seizure EEG signals in each specified frequency sub-band. Features applied in previous studies included the \emph{mean}, \emph{absolute mean}, \emph{variance}, \emph{SD}, \emph{skewness}, \emph{kurtosis}, \emph{absolute median}, and \emph{minimum and maximum values}.
	\item The \emph{energy}, \emph{average power}, and \emph{RMS} are used to observe the amplitudes of signals corresponding to specific frequency sub-bands.
	\item The \emph{line length} is used to estimate the fractal dimension of the DWT decomposition coefficients.
	\item The \emph{ShEn, ApEn, PE}, and \emph{WPE}, entropy-based indicators, determine the uncertainties and complexities of decomposed signals.
	\item Additionally, \emph{HE} and $\mathit{D_H}$ are applied to demonstrate randomness in a time series corresponding to the DWT coefficients of each sub-band. 
	\item \emph{Autoregressive} (\emph{AR}) and \emph{autoregressive-moving-average} (\emph{ARMA}) \emph{model parameters} of a time series model are fitted to DWT coefficient sequences. It is assumed that the time-series models explaining the dynamics of ictal and normal EEG patterns are different. Time-series model parameters are then estimated for each sub-band where stationarity assumption holds. However, it is claimed that ARMA parameters were significant features only in some sub-bands~\cite{gupta2018novel}.
\end{enumerate}

\subsection{Complexity of feature extraction}
Suppose a raw EEG time series is split into epochs; each of which contains $N$ time samples that are extracted into a feature. Based on mathematical expressions of each feature, we describe its computational complexity as a function of $N$ samples. Firstly, simple features that require only $\mathcal{O}(N)$ are standard statistical parameters, except those which need a prior sorting process, including  mode, median, $\Qf$, $\Qt$, and IQR, which require $\mathcal{O}(N\log N)$, energy, NE, RMS, line length, Hjorth parameters, number of zero-crossing, and number of local extrema. The HE, DFA, and fractal dimension features involve calculating a partial time series and solving a regression with one variable, so the first task is the more dominant. The complexity of these features is still linear in $N$, $\mathcal{O}(N)$. Similarly, the main calculation of LBP, LNDP, and LGP are from computing partial time series of length $m$. Since the computational complexity of computing each partial time series is linear in $m$ and there exists $N-m$ partial time series where $m<<N$, then the total complexity for calculation of these features is $\mathcal{O}(N)$, or $\mathcal{O}(Nm)$ (to be more precise).

Complexities of entropy-related features differ in the methods of approximating the probability distribution (a bubble sort algorithm requires $\mathcal{O}(N^2)$, while a quicksort takes $\mathcal{O}(N\log N)$~\cite{cormen2009introduction}). For exmaple, the ShEn needs $\mathcal{O}(N\log N)$. Also, the ApEn, FuzzyEn, DistEn, and SampEn normally require complexities of $\mathcal{O}(N^2)$ because there are approximately $N(N-1)/2$ similarity comparisons between two template vectors~\cite{richman2000physiological,manis2008fast}. Additionally, some complexities of entropy calculation can be reduced to $\mathcal{O}(N\log N)$ using a sorting algorithm before computing the self-similarity~\cite{fele2013faster,manis2018low}. Moreover, the PE and WPE require comparison of $m!$ permutation patterns and a template vector of length $m$. So there exist $(m!)(N-m+1)$ comparisons, making the computational complexity $\mathcal{O}((m!)N)$. On the other hand, the SVDEn needs SVD computation at the beginning followed by the entropy calculation. The complexity of calculating SVD is $\mathcal{O}(m^2N+m^3)$, where $m$ is the number of the singular values and $m<N$ and the entropy calculation requires $\mathcal{O}(m)$. Therefore, the complexity of SVDEn is dominated by the SVD computation of $\mathcal{O}(m^2N+m^3)$.

Computing FDFs firstly requires the Fourier transform that needs $\mathcal{O}(N\log N)$ for the fast-Fourier transform. Subsequent processes in calculating energy, IWMF, IWBW, SEF are just linear in $N$, so the complexity of the FDFs is $\mathcal{O}(N\log N)$. Computational complexities of TFDFs can be dominated by two processes: signal decomposition and feature extraction, depending on which step has a higher complexity. For example, the DWT has the complexity of $\mathcal{O}(N)$, while the STFT has the complexity of $\mathcal{O}(N \log m)$, where $m$ is a size of the moving window. If the energy feature is calculated from STFT, then the overall complexity is $\mathcal{O}(N \log m)$. On the other hand, the total complexity is $\mathcal{O}(N \log N)$ when the ApEn is calculated from DWT coefficients.

\section{Epileptic seizure detection}
\label{sec:use_feature}
\subsection{Performance metrics}
The method to assess AESD is critical for comparing detection algorithms. As a binary classification, seizure detection results from a classifier are comprised of true positive (TP), false positive (FP), false negative (FN), and true negative (TN). These quantities can be formed into a true positive rate (TPR) and a false positive rate (FPR), or as sensitivity (Sen), specificity (Spec) and accuracy (Acc). The accuracy is the overall classification performance, where the sensitivity and specificity are regarded as how well the algorithm detects seizure and normal outcomes, respectively. Currently, there are two sets of evaluation metrics in the AESD: epoch-based and event-based metrics~\cite{temko2011performance}. 

Epoch-based metrics are indicators used to evaluate the detection performance when each epoch segmented from a long EEG signal is supposed to be a separate data sample. The epoch-based metrics are calculated from the numbers of TP, FP, FN, and TN evaluated on all samples. For example, accuracy, sensitivity, and specificity have been reported in many studies~\cite{acharya2013automated,guo2010automatic,aarabi2006automated}. Nevertheless, these epoch-based metrics can lead to wrong diagnosis when seizure events are incorrectly recognized.
For instance, an algorithm wrongly classifying one totally short seizure event as normal still achieves a high epoch-based performance since there are many other epochs that are correctly detected.

Event-based metrics, on the other hand, are used to evaluate a classifier based on seizure events in EEG signals.
Two common metrics, good detection rate (GDR) and FPR per hour (FPR/h), are calculated based on the intersection of detection results and annotations~\cite{vidyaratne2017real,shoeb2010application,satirasethawong2015amplitude}. Here, GDR is defined as the percentage of detected seizure events that have an overlap with the annotations and FPR/h is the proportion of events declared as a seizure without any intersection with the annotations in one hour. A higher GDR indicates a higher number of correctly detected seizure events, while a small FPR/h refers to having a lower number of wrongly recognized seizure events. However, care is required with these high event-based metrics to avoid being misled into a conclusion of a correct detection when a duration is considered. For example, declaring an occurrence of seizure at the last second of an actual seizure event is still counted as good detection even though the detection system nearly misses the whole seizure event.

\subsection{Selected features used in epileptic seizure detection}
From the lists of common features described in Section~\ref{sec:feature_extraction}, this section summarizes how those features were applied to AESD in the literature. Performances of the detection algorithms generally depend on both feature selection and the classifier, so we report the classification method, features and categories by domains, and the performance results in Tables~\ref{tab:single_domain_chb} and \ref{tab:multi_domain_chb} for studies using the CHB-MIT Scalp EEG database~\cite{Goldbergere215} and in Tables~\ref{tab:single_domain_bonn} and \ref{tab:multi_domain_bonn} for those using the Bonn University database~\cite{andrzejak2001indications}. The description of the CHB-MIT Scalp EEG database is explained in \Cref{sec:exp}. For the Bonn University database, there are five sets of data, each of which contains 100 single-channel EEG samples. Only one data set consisted of seizure activities recorded intracranially from epileptic patients, whereas the others were regarded as non-epileptic EEG signals. More data description is available in~\cite{andrzejak2001indications}. We have separated these summaries according to the databases so that a fair comparison of performance results can be concluded. However, even though these studies tested the methods on the same database, the selection of data records in each paper can be varied or this information was not available. In this case, the interpretation of performance figures should be performed with care, and we have put a footnote to such publications.
The methods are described as a series of transformations or pre-processing steps (if performed) followed by classifiers.

\begin{table}[h!]
	\caption{Summary of automated epileptic seizure detection using the \emph{CHB-MIT} Scalp EEG database when \emph{single-domain} features were used.}
	\centering
		\resizebox{1\linewidth}{!}{
		\begin{tabular}{l p{0.58\linewidth} lll}
\toprule

Domain         & Features                                                                                                   & Method                  & Performance              & Ref.                         \\
\toprule
Time           & Raw signal                                                                                                 & ANN                     & Acc = 100\%              & \cite{chakraborti2018machine}       \\
				& 
				EEG                                                                                                       & Thresholding            & Sen = 88.50\%, FPh = 0.18 & \cite{satirasethawong2015amplitude} \\
				& Line length, NE, variance, average power, max
& RBF SVM					 	& Acc = 95.17\%, Sen = 66.35\%, Spec = 96.91\% & \cite{janjarasjitt2017epileptic} \\
               & Absolute mean values, average power, SD, ratio of absolute mean values, skewness, kurtosis & MSPCA + EMD + RF   & Acc = 96.90\%             & \cite{alickovic2018performance}     \\
               &                                                                                                           
& MSPCA + EMD + SVM  & Acc = 97.50\%             & \cite{alickovic2018performance}     \\
               &                                                                                                            & MSPCA + EMD + ANN  & Acc = 96.90\%             & \cite{alickovic2018performance}     \\
               &                                                                                                            & MSPCA + EMD + k-NN & Acc = 94.90\%             & \cite{alickovic2018performance}     \\
               & $\DB$ & RVM          & Sen = 97.00\%, FPR/h = 0.24  & \cite{vidyaratne2017real}           \\
\midrule
Frequency      & Energy                                                                                                     & RBF SVM                 & Sen = 96.00\%, FPR/h = 0.08   & \cite{shoeb2010application}$^*$         \\
\midrule
Time-frequency 	& Spectrogram                                                                                                & STFT + SSDA               & Acc = 93.82\%            & \cite{yuan2017multi}               \\
& Mean, ratio of variance, SD, skewness, kurtosis, mean frequency, peak frequency            & DWT + ELM                 & Acc = 94.83\%            & \cite{ammar2016seizure}         \\
				& Log of variance
& DWT + thresholding		& Acc = 93.24\%, Sen = 83.34\%, Spec = 93.53\% & \cite{janjarasjitt2017performance} \\
& & DWT + RBF SVM			& Acc = 96.87\%, Sen = 72.99\%, Spec = 98.13\% & \cite{janjarasjitt2017epileptic} \\
			   & Absolute mean, average power, SD, ratio of absolute mean, skewness, kurtosis & MSPCA$^1$ + DWT + RF     & Acc = 100\%              & \cite{alickovic2018performance}     	\\
               &                                                                                                            & MSPCA + DWT + SVM    & Acc = 100\%              & \cite{alickovic2018performance}     \\
               &                                                                                                            & MSPCA + DWT + ANN    & Acc = 100\%              & \cite{alickovic2018performance}     \\
               &                                                                                                            & MSPCA + DWT + k-NN   & Acc = 100\%              & \cite{alickovic2018performance}     \\
               &                                                                                                            & MSPCA + WPD$^2$ + RF     & Acc = 100\%              & \cite{alickovic2018performance}     \\
               &                                                                                                            & MSPCA + WPD + SVM    & Acc = 100\%              & \cite{alickovic2018performance}     \\
               &                                                                                                            & MSPCA + WPD + ANN    & Acc = 100\%              & \cite{alickovic2018performance}     \\
               &                                                                                                            & MSPCA + WPD + k-NN   & Acc = 100\%              & \cite{alickovic2018performance}     \\
               & Energy                                                                                                     & HWPT$^3$ + RVM                & Sen = 97.00\%, FPR/h = 0.25  & \cite{vidyaratne2017real}           \\
               & Energy, $\DB$                                                                                  & HWPT + RVM                & Sen = 97.00\%, FPR/h = 0.10    & \cite{vidyaratne2017real}           \\
\bottomrule 
\multicolumn{5}{l}{$^1$ Multi-scale principal component analysis, $^2$ wavelet packet decomposition,$^3$ harmonic wavelet packet transform} \\
$^*$ Use all data records  & & & 
\end{tabular}
}
	\label{tab:single_domain_chb}
\end{table}

\begin{table}[h!]
	\caption{Summary of automated epileptic seizure detection using the \emph{CHB-MIT} Scalp EEG database when \emph{multi-domain} features were used.}
	\centering
		\resizebox{1\linewidth}{!}{
		\begin{tabular}{p{0.3\linewidth} p{0.25\linewidth}llll}
		\toprule
Time                                              & Frequency                        & Time-frequency & Method                & Performance             &  Ref.           \\
\toprule
Variance, RMS, skewness, kurtosis, SampEn & Peak frequency, median frequency &                & LDA                   & Sen = 70.00\%, Spec = 83.00\% & \cite{fergus2016machine} \\
                                                  &                                  &                & QDA                   & Sen = 65.00\%, Spec = 92.00\% & \cite{fergus2016machine} \\
                                                  &                                  &                & Polynomial classifier & Sen = 70.00\%, Spec = 83.00\% & \cite{fergus2016machine} \\
                                                  &                                  &                & Logistic regression   & Sen = 79.00\%, Spec = 86.00\% & \cite{fergus2016machine} \\
                                                  &                                  &                & k-NN                  & Sen = 84.00\%, Spec = 85.00\% & \cite{fergus2016machine} \\
                                                  &                                  &                & DT                    & Sen = 78.00\%, Spec = 80.00\% & \cite{fergus2016machine} \\
                                                  &                                  &                & Parzen classifier     & Sen = 61.00\%, Spec = 86.00\% & \cite{fergus2016machine} \\
                                                  &                                  &                & SVM                   & Sen = 79.00\%, Spec = 86.00\% & \cite{fergus2016machine}\\
\bottomrule
\end{tabular}
        }
	\label{tab:multi_domain_chb}
\end{table}

\subsubsection{Time-domain features (TDFs)} Previous studies using TDFs in both databases are depicted in Tables~\ref{tab:single_domain_chb} and~\ref{tab:single_domain_bonn}. The most obvious approach was to use the raw EEG signals as inputs but a deep learning process to extract the patterns. Using the raw signal with ANN~\cite{chakraborti2018machine} and one-dimensional CNN following the $z$-score normalization process~\cite{acharya2018deep} provided promising results.
However, data used in~\cite{chakraborti2018machine} were only chosen from female patients, where the records contained epileptic activities in the frontal lobe and the activities were only simple and complex partial epilepsy.
A thresholding technique using aEEG was used in~\cite{satirasethawong2015amplitude}, where the threshold could be simultaneously updated. The proposed method effectively detected a high-amplitude seizure, but it also responded to high-amplitude artifacts. Thus, it is unsuitable for detecting a seizure when the EEG signals are contaminated with artifacts. Another disadvantage of the detection algorithm was that it required the EEG signal to be normal at the beginning and ictal patterns must be adequately long. Hence, only 85 seizure events were left for evaluation. When a single feature was extracted to detect epilepsy, amplitude or uncertainty-related features were commonly used with a ML technique. Results of using LBP with k-nearest neighbor (k-NN) showed that histograms between normal and epileptic patterns were different~\cite{kumar2015classification}.
Similarly, LNDP and LGP were individually applied to k-NN, SVM, decision tree (DT), and ANN to distinguish their patterns in epileptic EEG signals~\cite{jaiswal2017local}. As a result, both LNDP and LGP achieved capabilities of analyzing seizures for any classifiers while k-NN had the fastest computation. In addition, PE was employed to distinguish seizures from EEG signals where it was concluded that PE from a normal EEG signal was higher than that from a seizure EEG signal~\cite{li2014using}. Furthermore, WPE, which was calculated from sub-segments of one-long epoch and then concatenated into a feature vector, was used with three classifiers: linear SVM, radial basis function kernel SVM (RBF SVM), and ANN to find the performance of each classifier~\cite{tawfik2016hybrid}. Combining WPE and RBF SVM outperformed the other two classifiers. Unlike the ML approach, the results of using a thresholding method with ApEn as a single feature showed that the seizure period obtained lower ApEn than the normal period~\cite{ocak2009automatic}. An error energy obtained from a linear predictive filter was used in a thresholding method in~\cite{altunay2010epileptic}, where the epileptic group had a higher error energy than the normal group.

On the other hand, making use of several TDFs applied to ML techniques was found in~\cite{alickovic2018performance, li2018detection} (Tables~\ref{tab:single_domain_chb} and~\ref{tab:single_domain_bonn}). In~\cite{alickovic2018performance}, statistical features, including absolute mean, SD, skewness, kurtosis, ratio of absolute mean, and average power of each IMF obtained from EMD, were computed following an artifact removal using a multi-scale principal component analysis (MSPCA), and then passed to many classifiers, including SVM, ANN and k-NN.
Applying FuzzEn and DistEn in~\cite{li2018detection} to a multi-length segmentation approach and a quadratic discriminant analysis (QDA) showed that this combination could identify seizures. However, their performance was inferior to that in~\cite{jaiswal2017local} which used only single features with ANN or k-NN. Alternatively, among pioneer works in epilepsy detection (not listed in the tables as the data were from different sources), there were studies considering a group of features applied to a criterion-based classifier as opposed to ML approaches~\cite{gotman1982automatic,gotman1976automatic,dingle1993multistage}. The average amplitude relative to the background, average duration, and CV of durations of half-wave transformation were applied to detect epilepsy \cite{gotman1982automatic}. A half-wave  established in~\cite{gotman1976automatic} was proposed to determine the amplitude and the duration of the wave. The results showed that the CV should be small, indicating a measure of rhythmicity of the EEG signal, the average duration was limited in some specific range, and the average amplitude was high. Similarly, after the half-wave transformation, duration, amplitude, and sharpness (defined by the slope of the half-wave) were extracted and then compared to its background activity~\cite{dingle1993multistage}, which revealed that these features were in a specific range during epileptic seizures. However, the specific range was not provided in~\cite{dingle1993multistage}. Energies from three IMFs performed by a thresholding method from the epilepsy period were higher compared to the normal~\cite{orosco2009epileptic}.

\begin{table}[h!]
	\caption{Summary of automated epileptic seizure detection using the \emph{Bonn University} database when \emph{single-domain} features were used.}
	\centering
		\resizebox{1\linewidth}{!}{
		\begin{tabular}{l p{0.58\linewidth} lll}
\toprule

Domain & Features & Method & Performance & Ref. \\
\toprule
Time & Raw signal & CNN & Acc = 88.70\%, Sen = 95.00\%, Spec = 90.00\% & \cite{acharya2018deep} \\
 & Linear prediction error energy & Thresholding & Sen = 92.00\%, Spec = 94.50\% & \cite{altunay2010epileptic} \\
 & LBP & k-NN & Acc = 99.33\% & \cite{kumar2015classification} \\
 & LNDP & ANN & Acc = 98.72\%, Sen = 98.30\%, Spec =  98.82\% & \cite{jaiswal2017local} \\
 & 1D-LGP & ANN & Acc = 98.65\%, Sen = 98.44\%, Spec = 98.70\% & \cite{jaiswal2017local} \\
 & FuzzyEn, DistEn & QDA & Acc = 92.80\%, Sen = 90.67\%, Spec = 92.80\% & \cite{li2018detection} \\
 & ApEn & Thresholding & Acc = 73.00\% & \cite{ocak2009automatic} \\
 & WPE & Linear SVM & Acc = 91.63\% & \cite{tawfik2016hybrid} \\
 &  & RBF SVM & Acc = 93.38\% & \cite{tawfik2016hybrid} \\
 &  & ANN & Acc = 91.86\% & \cite{tawfik2016hybrid} \\
\midrule
Frequency & PSD & DT & Acc = 98.72\%, Sen = 99.40\%, Spec = 99.31\% & \cite{polat2007classification} \\
	& Peak amplitude, peak frequency & PSD (ARMA) + GMM & Sen = 70.00\%, Spec = 53.33\% & \cite{faust2010automatic} \\
 &  & PSD (ARMA) + ANN & Sen = 91.67\%, Spec = 83.33\% & \cite{faust2010automatic} \\
 &  & PSD (ARMA) + RBF SVM & Sen = 96.67\%, Spec = 86.67\% & \cite{faust2010automatic} \\
 &  & PSD (YW) + GMM & Sen = 78.33\%, Spec = 83.33\% & \cite{faust2010automatic} \\
 &  & PSD (YW) + ANN & Sen = 98.33\%, Spec = 90.00\% & \cite{faust2010automatic} \\
 &  & PSD (YW) + RBF SVM & Sen = 96.67\%, Spec = 86.67\% & \cite{faust2010automatic} \\
 &  & PSD (Burg) + GMM & Sen = 75.00\%, Spec = 93.33\% & \cite{faust2010automatic} \\
 &  & PSD (Burg) + ANN & Sen = 98.33\%, Spec = 96.67\% & \cite{faust2010automatic} \\
 &  & PSD (Burg) + RBF SVM & Sen = 98.33\%, Spec = 96.67\% & \cite{faust2010automatic} \\
\midrule
Time-frequency & log of Fourier spectrum at scale 4 and 5 & DT-CWT$^4$ + k-NN & Acc = 100\% & \cite{chen2014automatic} \\
 & Mean, variance, skewness, kurtosis, entropy & EWT$^5$ + RBF SVM & Acc = 100\% & \cite{anand2017automatic} \\
 & Absolute coefficients, absolute mean, absolute median, absolute SD, absolute max, absolute min & RDSTFT$^6$ + ANN & Acc = 99.80\%, Sen = 99.90\%, Spec = 99.60\% & \cite{samiee2015epileptic} \\
 & Absolute mean, average power, SD & DWT + PCA + RBF SVM & Acc = 98.75\%, Sen = 99.00\%, Spec = 98.50\% & \cite{subasi2010eeg} \\
 &  & DWT + LDA + RBF SVM & Acc = 100\%, Sen = 100\%, Spec = 100\% & \cite{subasi2010eeg} \\
 &  & DWT + ICA + RBF SVM & Acc = 99.50\%, Sen = 99.00\%, Spec = 100\% & \cite{subasi2010eeg} \\
 & Absolute max, absolute min, absolute mean, SD & WPD + k-NN & Acc = 99.45\% & \cite{wang2011best} \\
 & Line length & DWT + ANN & Acc = 97.77\% & \cite{guo2010automatic} \\
 & Fractional energy & SPWVD$^7$ + ANN & Acc = 99.92\% & \cite{tzallas2007automatic} \\
 & ApEn & MWT$^8$ (GHM) + ANN & Acc = 96.69\%, Sen = 98.62\%, Spec = 89.91\% & \cite{guo2010epileptic} \\
 &  & MWT (CL) + ANN & Acc = 95.15\%, Sen = 96.57\%, Spec = 89.21\% & \cite{guo2010epileptic} \\
 &  & MWT (SA4) + ANN & Acc = 98.27\%, Sen = 99.00\%, Spec = 95.50\% & \cite{guo2010epileptic} \\
 &  & DWT + Thresholding & Acc = 96.00\% & \cite{ocak2009automatic} \\
 & WPE & DWT + Linear SVM & Acc = 86.50\% & \cite{tawfik2016hybrid} \\
 &  & DWT + RBF SVM & Acc = 88.25\% & \cite{tawfik2016hybrid} \\
 &  & DWT + ANN & Acc = 86.63\% & \cite{tawfik2016hybrid} \\
 & HE, ARMA parameters & DCT + RBF SVM & Acc = 97.79\%, Sen = 97.97\%, Spec = 97.60\% & \cite{gupta2018novel} \\
 & HE, fractal dimension, PE & DT-CWT + RBF SVM & Acc = 98.87\%, Sen = 98.20\%, Spec = 100\% & \cite{li2017automatic} \\
 &  & DT-CWT + k-NN & Acc = 97.80\%, Sen = 97.20\%, Spec = 98.60\% & \cite{li2017automatic} \\
 &  & DT-CWT + DT & Acc = 90.33\%, Sen = 90.00\%, Spec = 94.00\% & \cite{li2017automatic} \\
 &  & DT-CWT + RF & Acc = 98.13\%, Sen = 98.20\%, Spec = 98.40\% & \cite{li2017automatic} \\
 
\bottomrule 
\multicolumn{5}{l}{$^4$ Dual-tree complex wavelet transform, $^5$ empirical wavelet transform, $^6$ rational discrete STFT, $^7$ smoothed pseudo Wigner-Ville distribution, $^8$ multiwavelet transform} \\
\multicolumn{5}{l}{Note that all publications used every set of the data but none of them explained which segments were selected.}

\end{tabular}
}
	\label{tab:single_domain_bonn}
\end{table}

\begin{table}[h!]
	\caption{Summary of automated epileptic seizure detection using the \emph{Bonn University} database when \emph{multi-domain} features were used.}
	\centering
		\resizebox{1\linewidth}{!}{
		\begin{tabular}{p{0.3\linewidth} p{0.12\linewidth} p{0.3\linewidth}lll}
		\toprule
Time & Frequency & Time-frequency & Method & Performance & Ref. \\
\toprule
Mean, energy, SD, max value &  & Mean, energy, SD, max value & DWT + HHT + NNE & Acc = 98.78\% & \cite{li2017classification} \\
 &  &  & DWT + HHT + ANN & Acc = 88.00\% & \cite{li2017classification} \\
 &  &  & DWT + HHT + RNN & Acc = 91.33\% & \cite{li2017classification} \\
 &  &  & DWT + HHT + SVM & Acc = 94.67\% & \cite{li2017classification} \\
 &  &  & DWT + HHT + k-NN & Acc = 95.33\% & \cite{li2017classification} \\
 &  &  & DWT + HHT + LDA & Acc = 92.67\% & \cite{li2017classification} \\
Max, min, mean, SD, kurtosis, skewness, Q1, Q3, IQR, median, mode, mobility, complexity, SampEn, HE, DFA &  & Max, min, mean, SD, ShEn & DWT + RF & Acc = 97.40\%, Sen = 97.40\%, Spec = 97.50\% & \cite{mursalin2017automated} \\
$\DB$ &  & Energy & HWPT + RVM & Acc = 99.80\%, Sen = 100\%, Spec = 99.00\% & \cite{vidyaratne2017real}\\
\bottomrule
\multicolumn{5}{l}{Note that all publications used every set of the data but none of them explained which segments were selected.}
\end{tabular}
        }
	\label{tab:multi_domain_bonn}
\end{table}

\subsubsection{Frequency-domain features (FDFs)}
From Tables~\ref{tab:single_domain_chb} and~\ref{tab:single_domain_bonn}, commonly-used features in the  frequency domain were PSD, peak amplitude, peak frequency, and energy. In~\cite{polat2007classification}, the combination of PSD based on the Welch method and DT as a classifier achieved a high performance, both in terms of sensitivity and specificity. The first four local extrema and their frequencies obtained from the PSD of the signal computed by several PSD estimation methods (ARMA, Yule-Walker equation, and Burg) were used to identify epileptiform activities~\cite{faust2010automatic}. These features computed from any estimated PSD performed well with ANN and RBF SVM but the performance dropped when using these features with a Gaussian mixture model (GMM). By means of filter bank analysis, the energy obtained from each frequency range of three consecutive epochs was concatenated into a feature vector for temporal and spectral information, and then applied to RBF SVM for classification \cite{shoeb2010application}. As a result, the energies from the seizure group tended to be higher than that of the normal group. 

Furthermore, some features in the frequency domain were used jointly to detect epilepsy.
Both criteria-based and ML methods were used in the AESD.
The peak frequency, bandwidth of the peak frequency, and power ratio of the current and background epochs in the same frequency range were applied to the AESD in newborns \cite{gotman1997automatic}.
These features with appropriate thresholds could identify a large part of seizures but still incorrectly classified some seizure events, such as slow or spike waves.

\subsubsection{Time-frequency-domain features (TFDFs)}
Common transformed signals still containing information of both time and frequency domains were analyzed by STFT or DWT. Firstly, we describe the literature that regards the transformed signals as raw features. By using a deep learning approach, the STFT spectrogram of a raw EEG signal was considered as a feature with a modified stacked sparse denoising autoencoder (mSSDA) as a classifier~\cite{yuan2017multi}.
This combination was then compared to the other classifiers, where the result showed that the mSSDA could successfully distinguish epilepsy from normality.
Similarly, the logarithms of the Fourier spectrum of dual-tree complex transform (DT-CWT) coefficients at scales 4 and 5 were performed with k-NN for detecting epilepsy~\cite{chen2014automatic}. This approach performed well with an accuracy of 100\% and could be implemented in real time because of its low computational complexity, consuming 14.4 milliseconds for processing.

The other group of previous studies further extracted features from STFT or DWT sequences. 
Amplitude and uncertainty measurements were commonly used individually, whereas statistical parameters were jointly employed in general. Secondly, we describe studies that applied a single feature on transformed signals. 
A logarithm of variance of each DWT sub-band coefficients from a chosen channel were used independently with a threshold~\cite{janjarasjitt2017performance} to detect patient-specific epileptic seizure. 
The use of this feature obtained outstanding average performances of 93.24\% accuracy, 83.34\% sensitivity, and 93.53\% specificity from the best classification result from each chosen patient.
The author also applied combinations of those features to SVM and thoroughly compared the results with TDFs~\cite{janjarasjitt2017epileptic}.
The TDFs were line length, NE, variance, power, and maximum value of raw EEG signals.
As a result, the best average detection performances from each patient using SVM with wavelet-based features were average accuracy of 96.87\%, sensitivity of 72.99\%, and specificity of 98.13\%.
On the other hand, the best average outcomes using TDFs were 95.17\% accuracy, 66.35\% sensitivity, and 96.91\% specificity, respectively.
The energy calculated from each transformation was commonly used in seizure detection.
The energy from each DWT sub-band coefficients with RBF SVM was used to present the spectrum in each sub-band~\cite{shoeb2004patient}.
Similarly, fractional energies, energies in specific frequency bands and time windows of a Smoothed pseudo WVD (SPWVD) were applied to principal component analysis (PCA) and ANN, resulting in a high average accuracy of 95.00\% ~\cite{tzallas2007automatic}.

Furthermore, energies from Cohen's class transformation, including SPWVD, were employed to analyze EEG signals~\cite{tzallas2009epileptic}.
As a result, the energy computed from any transformation was successfully able to distinguish epileptic EEG from normal EEG.
The line length calculated from DWT coefficients with ANN could determine seizures properly~\cite{guo2010automatic}.
For the uncertainty-related features, the ApEn, calculated from DWT coefficients, was also applied to a criteria-based detection system~\cite{ocak2009automatic} and the ApEn computed from multiwavelet transform (MWT) was used with ANN~\cite{guo2010epileptic} in the AESD.
Three famous multiwavelets, namely Gernoimo-Hardin-Massopust (GHM), Chui-Lian (CL), and SA4, were exploited.
Both approaches achieved an accuracy of 95-98\%.
In addition, as a result in~\cite{ocak2009automatic}, the ApEn had small values in a seizure group and using DWT could improve and provide some useful information for the AESD. 
When WPE was extracted from the DWT coefficients in~\cite{tawfik2016hybrid}, the work tested the methods of ANN, linear SVM, and RBF SVM, and the results indicated that RBF SVM could surpass the two other classifiers, but with lower accuracies than those of~\cite{guo2010epileptic} evaluated on the same data set.

Thirdly, we discuss TFDFs that were applied jointly. In this case, statistical parameters and amplitude-related measures were commonly used.
The absolute mean, SD, average power of DWT coefficients of each level, and ratio of absolute mean values of the adjacent sub-bands were combined and applied to the dimensionality reduction techniques of PCA, independent component analysis (ICA), and linear discriminant analysis (LDA) to extract useful features and reject meaningless features~\cite{subasi2010eeg}.
The dimension-reduced feature vector was classified by RBF SVM and the results showed that this feature-based approach could firmly detect seizure with an accuracy of 98-100\%.
The mean, SD, skewness, kurtosis, and ratio of variance of the adjacent sub-bands computed from the DWT coefficients of every decomposition level, and mean frequency and peak frequency calculated from the PSD of each decomposition level were cooperated with an extreme learning machine (ELM) to classify an epileptic EEG signal~\cite{ammar2016seizure}.
Due to the efficiency of computation in the ELM~\cite{huang2006extreme}, this combination could perform the detection accurately and quickly in a short time. Nevertheless, a total of 20 recordings from three patients was used to evaluate the classification performance.
The absolute mean, SD, skewness, kurtosis, ratio of absolute mean, and average power of each decomposition level were applied with DWT and wavelet packet decomposition (WPD) in the AESD~\cite{alickovic2018performance}.
A noise reduction of the multi-channel EEG signals was done by the MSPCA and when using this with the RF, SVM, ANN, and k-NN classifiers, it achieved an accuracy of 100\%. However, the authors only used  2,000 8-second EEG segments, 1,000 segments for each class, as samples in the experiments. Moreover, the mean, variance, skewness, kurtosis, and ShEn of each sub-band in the empirical wavelet transform (EWT) were extracted and the use of these features could potentially detect focal epilepsy automatically by RBF SVM with a reported accuracy of 100\%~\cite{anand2017automatic}.
The absolute mean, absolute median, absolute maximum, absolute minimum, and absolute values of coefficients from Rational discrete STFT (RDSTFT) were investigated with diverse classifiers~\cite{samiee2015epileptic}.
The results indicated that the ANN was the most optimal classifier, and the absolute median and some absolute coefficients were the most dominant features.

In addition to statistical parameters, combinations of uncertainty and similarity measurements were also established to indicate a periodic pattern of seizure.
The PE, HE, and $\Dh$ extracted from some DT-CWT sub-band coefficients were used as the features for RBF SVM, k-NN, DT and RF analyses~ \cite{li2017automatic}. These features from each decomposition level could separate the seizure activities from the normal group.
The HE estimated from discrete-time fractional Brownian motion process and discrete-time fractional Gaussian motion process and ARMA parameters from multi-rate filter bank were computed and applied to the RBF SVM to detect seizure \cite{gupta2018novel}.
As a result, the HE of each sub-band was different according to the stationarity of each sub-band, and the HE and the ARMA parameters computed from some sub-band coefficients were significant.

\subsubsection{Multi-domain features}
In terms of the complementary information that cannot be achieved in one domain, some studies have used features from different domains combined into a feature vector.
Table~\ref{tab:multi_domain_bonn} shows that features from the time domain and time-frequency domain were commonly used together since the time-frequency domain also provides spectral information.
A combination of the mean, SD, energy, and maximum value from EEG signal and from the envelope spectrum of some DWT sub-bands were also formed as features and applied to a neural network ensemble (NNE)~\cite{li2017classification}.
The authors reported that features extracted from the envelope spectrum of the decomposition levels of the seizure group were normally higher than in the normal group~\cite{li2017classification}.
The mean, SD, maximum and minimum values of EEG signal combined with the energy ratio of each detail and approximation coefficients from the DWT were employed with an Ant Colony classifier~\cite{salem2014epileptic}. As a result, these features were typically higher in epileptic seizure periods, which agreed with previous studies. 

The mean, SD, skewness, kurtosis, maximum and minimum values, median, mode, $\Qf$, $\Qt$, IQR, mobility, complexity, HE, DFA, ShEn, and SampEn were computed from an EEG signal, and the mean, SD, maximum and minimum values were obtained from each sub-band of DWT \cite{mursalin2017automated}.
Subsequently, a novel feature selection, the improved correlation-based feature selection, was proposed to select significant features based on their correlation and then the selected features were classified by the RF.
The results showed that the feature dimension was extremely reduced and the remaining features could improve the detection system compared to features selected by the CFS.
The variance, skewness, kurtosis, RMS, and SampEn calculated from each sub-band and complete bandwidth based on the second order Butterworth filter, and peak frequency and median frequency computed from the PSD of the bands were fed to several feature selection techniques and classifiers to find the best combination~\cite{fergus2016machine}; see performance in Table~\ref{tab:multi_domain_chb}.
From the best results reported in the article, the top five uncorrelated features via LDA a with backward search, a feature selection method, in each region were mostly the RMSs of some frequency range.
Moreover, compared to several classifiers, k-NN performed the best with this approach.
However, the authors selected only records containing seizure activities to perform the experiments.
In \cite{vidyaratne2017real}, the $\DB$ from raw EEG signals and energies from each Harmonic wavelet packet transform (HWPT) sub-bands used with a relevance vector machine (RVM) could be successfully applied in real-time settings.

On the other hand, some literature also included features from three domains to obtain information from all different aspects.
Twenty-one features from different domains, which were RMS, NE, line length, number of zero-crossings, number of local extrema, three Hjorth parameters, ShEn, ApEn, SVDEn, and error from validating an AR model on other segment obtained from an EEG signal, IWMF, IWBW, SEF90, the total power, peak frequency and bandwidth of the peak frequency calculated from PSD, and energy extracted from a specific DWT sub-band that corresponded to 1.25-2.5 Hz, were tested with LDA to determine their abilities on neonatal seizure detection in the EEG signal \cite{greene2008comparison}.
It showed that all the features combined achieved the best outcome, although the RMS, number of local extrema, and line length were three most significant features.

Based on our review, the first conclusive point is that a raw EEG signal or coefficients of any transformation were typically combined with a deep learning technique. Most studies were in favor of a ML approach rather than a thresholding technique, since the threshold of each feature can vary with subjects and baselines of EEG signals in a normal period.
Secondly, no statistical parameter was used individually in an AESD. In fact, statistical parameters were always applied with other features that were related to amplitude or uncertainty measurement and with a ML technique. This implies that statistical parameters alone do not attain sufficient information to distinguish the seizures from the normality. On the other hand, features, such as quantities of amplitude and similarity, were commonly used separately from other features. In particular, the energy was the most widely used feature that contributed to the meaning of amplitude. Moreover, attributes of amplitude and uncertainty were also typically considered when being employed with statistical parameters. This indicates that the amplitude-related features and similarity-uncertainty measurements have the capability to distinguish the seizures from the normality in an EEG signal.

\section{Methods for feature evaluation}
\label{sec:method_eva}
Our paper aims to examine the significance of a single feature to the classification performance through the Bayesian method. However, multiple features are commonly applied in practice, so a redundancy analysis is also needed. Therefore, this section describes the brief concepts about the Bayes classifier and the correlation-based feature selection.
\subsection{Bayes error rate}
Common feature selection and dimensionality reduction methods in AESD~\cite{aarabi2006automated, tzallas2007automatic,bandarabadi2015epileptic} do not provide information about how a single feature independently improves the classification performances. Hence, to determine the significance of each feature on the AESD, we employed the Bayes classifier. If we define $x$ as a feature, $C(x)$ as a class in which feature $x$ is classified, and $C_i$ denotes a class $i$ labeled from the data, then a misclassification error is generally defined as 
\begin{equation}
\mathrm{err} = \int\limits \sum\limits_{C_i\neq C(x)}P\left(C_i|x\right)p(x)dx,
\label{eq:error_general}
\end{equation}
where $P\left(C_i|x\right)$ is the posterior probability of $x$ in class $i$, and $p(x)$ is the probability density function of $x$.
Intuitively, we can interpret that the error is the total joint probability that the feature is incorrectly classified.
The Bayes optimal classifier gives the minimum error (the \emph{Bayes error rate}) by choosing the class of which the posterior probability is the highest~\cite{devroye2013probabilistic}.
As a result, the Bayes error rate ($\mathrm{err_{b}}$) is obtained from
\begin{equation}
\mathrm{err_{b}} = \int\sum\limits_{C_i\neq C_{\max}} P\left(C_i|x\right)p(x)dx,
\label{eq:bayes_general}
\end{equation}
where $C_{\max}$ is the class of which the posterior probability is maximum.

Practically, the distribution of a likelihood function is unknown, so a non-parametric distribution estimation was exploited in our experiment and the posterior probability was obtained by Bayes' rule: 
\begin{equation}
P\left(C_i|x\right) =\frac{p(x|C_i)P(C_i)}{p(x)},
\label{eq:bayesian_theorm}
\end{equation}
where $p(x|C_i)$ is the likelihood function, $P(C_i)$ is the prior distribution, and $p(x)$ is the evidence.
For all real $x$ values, the likelihood function can be estimated using the non-parametric kernel smooth function
\begin{equation*}
p\left(x|C_i\right) = \frac{1}{N_ih} \sum\limits_{j=1}^{N_i} K\left(\frac{x-x_j}{h}\right),
\label{eq:kernel_estimation}
\end{equation*}
where $N_i$ is the sample size of class $C_i$, $x_j$ is a sample in the class, $K(x)$ is the Gaussian kernel and $h$ is a bandwidth~\cite{fukunaga2013introduction,parzen1962estimation}.
Regarding a classification problem, the prior $P\left(C_i\right)$ is assumed to be binomial estimated by the size of class $C_i$ divided by the total number of samples. Finally, $p(x)$ is obtained by the total probability and it completes the calculation of $P\left(C_i|x\right)$ in \eqref{eq:bayesian_theorm}.

Our problem was a two-class classification (normal/seizure) problem, so the Bayesian error was reduced to
\begin{equation}
\mathrm{err_{b}} = \int \min_{i = 1,2} P\left(C_i|x\right)p(x)dx,
\label{eq:bayes_two}
\end{equation}
where $C_1$ and $C_2$ stand for the seizure and normal classes.
For the non-parametric kernel function, $h\approx 1.06\hat{\sigma} N_i^{-\frac{1}{5}} $ was chosen to be the optimal bandwidth, where $\hat{\sigma}$ is the sample SD of the sample.

We have observed that using conventional performance metrics, such as accuracy, cannot give the significance of each individual feature.
Hence, to evaluate the performance of individual features, we proposed to use an improvement rate (\textbf{rate}) from a standard condition ($\mathrm{err_0}$) as follows:
\begin{equation}
\mathrm{err_0} = \int\limits_{-\infty}^{\infty} P\left(C_2|x\right)p(x)dx = P\left(C_2\right),\quad
\mathbf{rate} =\frac{\mathrm{err_0} - \mathrm{err_b}}{\mathrm{err_0}} \times 100\%.
\label{eq:err0_imp}
\end{equation}
\subsection{Correlation-based feature selection (CFS)}
The CFS introduced in~\cite{hall1997feature} is a feature selection method based on the hypothesis that \emph{a good subset of features is highly correlated with the class, but uncorrelated with others}.
For a feature subset $F$ containing $k$ features, an index called heuristic merit is exploited to measure feature-feature and feature-class correlations and is defined by
\begin{equation}
	\mathrm{Merit}_F = \frac{k \bar{r}_{fc}}{\sqrt{k+k(k-1)\bar{r}_{ff}}},
	\label{eq:merit}
\end{equation}
where $\bar{r}_{fc}$ is the mean value of the correlation of feature and class, and $\bar{r}_{ff}$ is the average of the feature-feature correlation.
To find the subset $F$ with the highest merit score, we applied the CFS algorithm provided in~\cite{zhao2010advancing}, where the correlations are estimated by conditional entropy.
The algorithm initially assigns the subset $F$ to be empty.
New feature subsets are constructed by adding another feature that has not been previously selected to $F$.
All new subsets are then evaluated and the subset having the best merit score is used as the subset $F$ in the next iteration.
This process is repeated until $F$ has $m$ features.
The final subset $F$ contains features ranked in the descending order by the merit score.

\section{Experimental results}
\label{sec:exp}
The experiment was performed on the public CHB-MIT Scalp EEG database~\cite{Goldbergere215} that contains 24 EEG recordings from 23 patients: five males aged 3-22 years, 17 females aged 1.5-19 years, and one anonymous subject.
All EEG data were collected at the Children's Hospital Boston and the international 10-20 system was used to locate electrode placements.
All signals were recorded with a sampling frequency of 256 Hz digitized at 16-bit resolution and stored in an EDF file~\cite{shoeb2009application}.
This full database is publicly downloaded at PhysioNet (\url{https://physionet.org/physiobank/database/chbmit/}).
\begin{table}[h!]
        \caption{List of all records used to evaluate features.}
        \centering
		\resizebox{1\linewidth}{!}{
		\begin{tabular}{cccccccccc}
            \hline
            \multicolumn{8}{c}{\textbf{Records}} \\
            \hline
            chb01\_04  & chb01\_16  & chb02\_16+ & chb02\_19 & chb03\_03 & chb03\_35 & chb04\_08 & chb04\_28 \\
            chb05\_06  & chb05\_13  & chb06\_01  & chb06\_04 & chb07\_13 & chb07\_19 & chb08\_02 & chb08\_05 \\
            chb09\_06  & chb09\_08  & chb10\_38  & chb10\_89 & chb11\_92 & chb11\_99 & chb12\_33 & chb12\_38 \\
            chb13\_19  & chb13\_55  & chb14\_04  & chb14\_18 & chb15\_06 & chb15\_15 & chb16\_17 & chb16\_18 \\
            chb17a\_04 & chb17b\_63 & chb18\_32  & chb18\_35 & chb19\_29 & chb19\_30 & chb20\_14 & chb20\_16 \\
            chb21\_21  & chb21\_22  & chb22\_21  & chb22\_22 & chb23\_06 & chb23\_09 & chb24\_04 & chb24\_11 \\
            \hline
        \end{tabular}
        }
		\label{tab:chb_record}
\end{table}

We randomly chose two records from each case, subject to the inclusion condition that every record must contain at least one seizure activity, so the records for assessing the improvement rate of the Bayes error rate of each feature are shown in \Cref{tab:chb_record}. The channels were sequentially listed as follows: FP1-F7, F7-T7, T7-P7, P7-O1, FP1-F3, F3-C3, C3-P3, P3-O1, FP2-F4, F4-C4, C4-P4, P4-O2, FP2-F8, F8-T8, T8-P8, and P8-O2.
\begin{table}[h!]
\renewcommand{\arraystretch}{1}
\centering
\caption{List of features for the Bayesian error rate evaluation and redundancy analysis.}
\resizebox{1\linewidth}{!}{
\begin{tabular}{p{0.13\linewidth} p{0.8\linewidth}}
	\toprule%
	Domains & Features \\
	\toprule
	Time & Mean, variance, CV, skewness, kurtosis, max, min, energy, NE, line length, ShEn, ApEn, SampEn, number of zero-crossing, number of local extrema, mobility, complexity \\
	\midrule
	Frequency & IWMF, IWBW, SE, peak frequency, peak amplitude \\
	\midrule
	Time-frequency & Mean, absolute mean, variance, skewness, kurtosis, max, min, energy, line length \\
	\bottomrule
\end{tabular}
}
\label{tab:feature_list}
\end{table}

An EEG epoch was defined by segmenting a raw EEG signal in every channel with a  $4$-second width.
The next consecutive epoch was segmented from the raw EEG signal by moving the window for 1 second.
These choices were selected from inspecting the processing step using the commercial Persyst~\cite{persyst2019seizure} software.
After the process of segmenting the raw EEG signals, a feature was then computed from each epoch and each channel independently. Only commonly used features in the literature were selected in this experiment (\Cref{tab:feature_list}). For the ApEn and SampEn, the template length, $m$, and the tolerance, $r$, were set to $m=2$ and $r=0.2\mathrm{SD}$, where $\mathrm{SD}$ was the sample SD of the segment. All TDFs were calculated from a raw EEG signal, whereas features from the frequency domain were extracted from PSD and TFDFs were computed from DWT coefficients with the Daubechies wavelet tap 4 for five levels.

The EEG channel selection is still an open research question and using multi-channel EEG signals may be redundant. Moreover, some commercial software also analyzes the seizure activity over the left and right sides of the brain. For these reasons, we have not explored the channel selection topic but rather used a spatial averaging of features over the left and right sides of the brain, denoted as $x_{\mathrm{left}}$ and $x_{\mathrm{right}}$, respectively. We then estimated the posterior probability distributions and computed the Bayes error rates of using the left and right feature representatives. 

\subsection{Feature significance}
\label{sec:feat_sig}

\begin{table}[t]
	\caption{Bayes error ($\mathrm{err_b}$) and improvement rate of time-domain and frequency-domain features.}
\label{tab:imp_t_f}
\vspace{3mm}
\begin{subtable}[t]{0.50\textwidth}
\centering
\caption{\textbf{Time-domain} features.}
\label{subtab:imp_t}
\resizebox{1\linewidth}{!}{
\begin{tabular}{lcccc}
	\toprule
	\multirow{2}{*}{Feature} & \multicolumn{2}{c}{Left side} & \multicolumn{2}{c}{Right side}\\
	\cline{2-5}
	& \textbf{$\mathrm{err_b}$} & \textbf{rate} & \textbf{$\mathrm{err_b}$} & \textbf{rate}\\
	\toprule
Mean                     & 0.0174 & 0.00 & 0.0174 & 0.05\\
\textbf{Variance}                 & 0.0160 & \textbf{8.20} & 0.0166 & \textbf{4.78}\\ 
CV						 & 0.0174 & 0.00 & 0.0174 & 0.00\\
Skewness                 & 0.0174 & 0.00 & 0.0174 & 0.00\\
Kurtosis                 & 0.0174 & 0.08 & 0.0174 & 0.17\\ 
Max                      & 0.0174 & 0.00 & 0.0174 & 0.00\\ 
Min                      & 0.0174 & 0.00 & 0.0174 & 0.00\\
\textbf{Energy}                   & 0.0160 & \textbf{8.18} & 0.0166 & \textbf{4.77}\\ 
\textbf{NE}         & 0.0157 & \textbf{10.07} & 0.0160 & \textbf{8.36}\\ 
Line length              & 0.0166 & 1.92 & 0.0167 & 1.16\\ 
\textbf{ShEn}          & 0.0174 & 0.33 & 0.0165 & \textbf{5.64}\\ 
ApEn      				  & 0.0174 & 0.00 & 0.0174 & 0.00\\
SampEn           		  & 0.0174 & 0.00 & 0.0174 & 0.00\\
Local extrema  		  & 0.0174 & 0.00 & 0.0174 & 0.00\\
Zero-crossing            & 0.0174 & 0.09 & 0.0174 & 0.09\\ 
Mobility                 & 0.0174 & 0.00 & 0.0174 & 0.00\\ 
Complexity               & 0.0174 & 0.00 & 0.0174 & 0.00\\
\bottomrule
\end{tabular}
}
\end{subtable}
\hfill
\begin{subtable}[t]{0.47\textwidth}
\centering
\caption{\textbf{Frequency-domain} features.}
\label{subtab:imp_f}
\resizebox{1\linewidth}{!}{
\begin{tabular}{lcccc}
	\toprule
	\multirow{2}{*}{Feature} & \multicolumn{2}{c}{Left side} & \multicolumn{2}{c}{Right side}\\
	\cline{2-5}
	& \textbf{$\mathrm{err_b}$} & \textbf{rate} & \textbf{$\mathrm{err_b}$} & \textbf{rate} \\
	\toprule
IWMF             & 0.0174 & 0.00 & 0.0174 & 0.00\\ 
IWBW             & 0.0174 & 0.00 & 0.0174 & 0.00\\ 
SE & 0.0174 & 0.00 & 0.0174 & 0.00\\ 
Peak amplitude   & 0.0174 & 0.00 & 0.0174 & 0.00\\ 
Peak frequency   & 0.0174 & 0.00 & 0.0174 & 0.00\\ 
\bottomrule
\end{tabular}
}
\end{subtable}
\end{table}

\begin{figure}[h!]
\centering
	\begin{subfigure}[b]{0.49\textwidth}
	\centering
         \includegraphics[width=\textwidth]{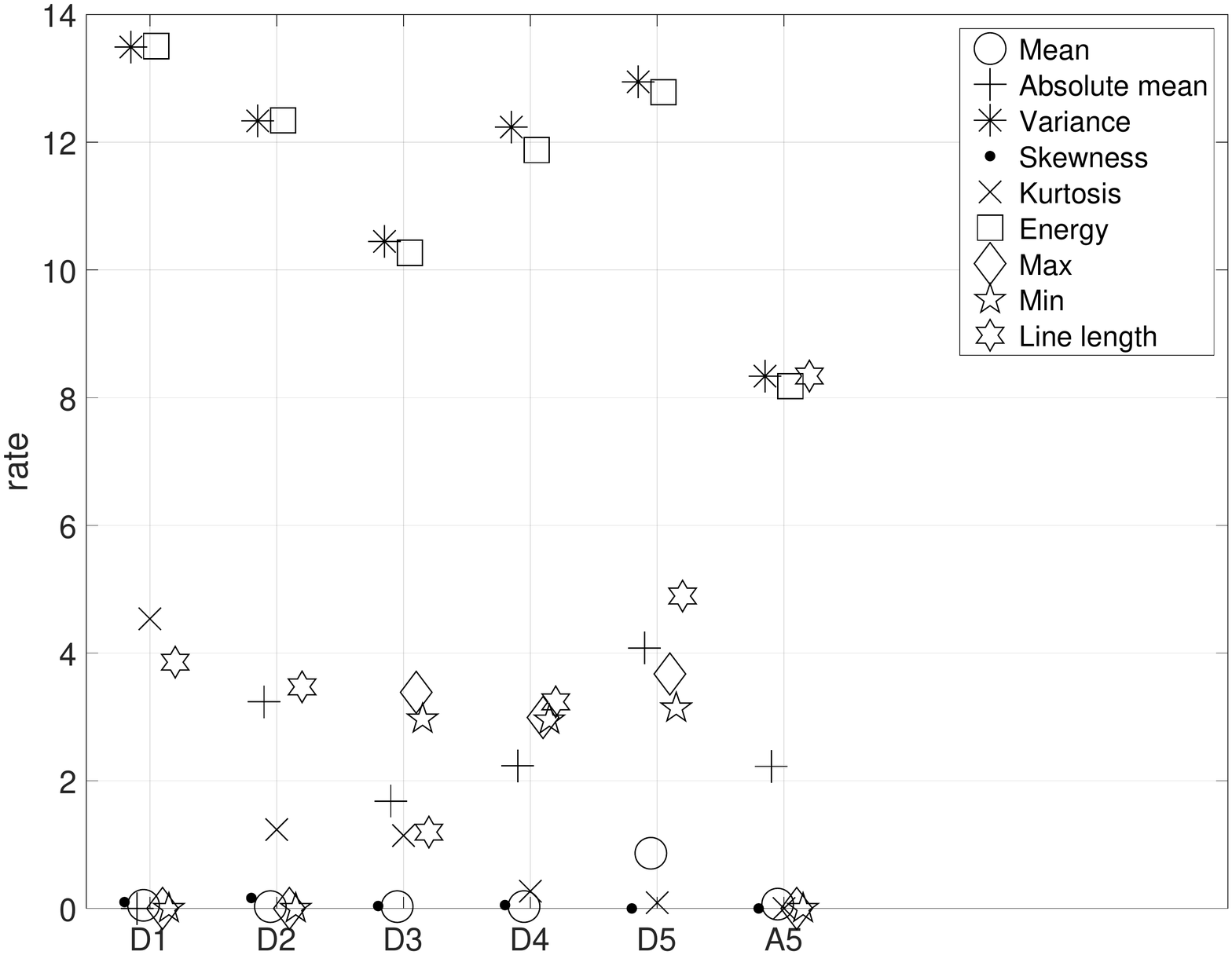}
         \caption{Improvement rate calculated from the \emph{left} hemisphere.}
         \label{subfig:imp_left_tf}
    \end{subfigure}
    \hfill
	\begin{subfigure}[b]{0.49\textwidth}
	\centering
    \includegraphics[width=\textwidth]{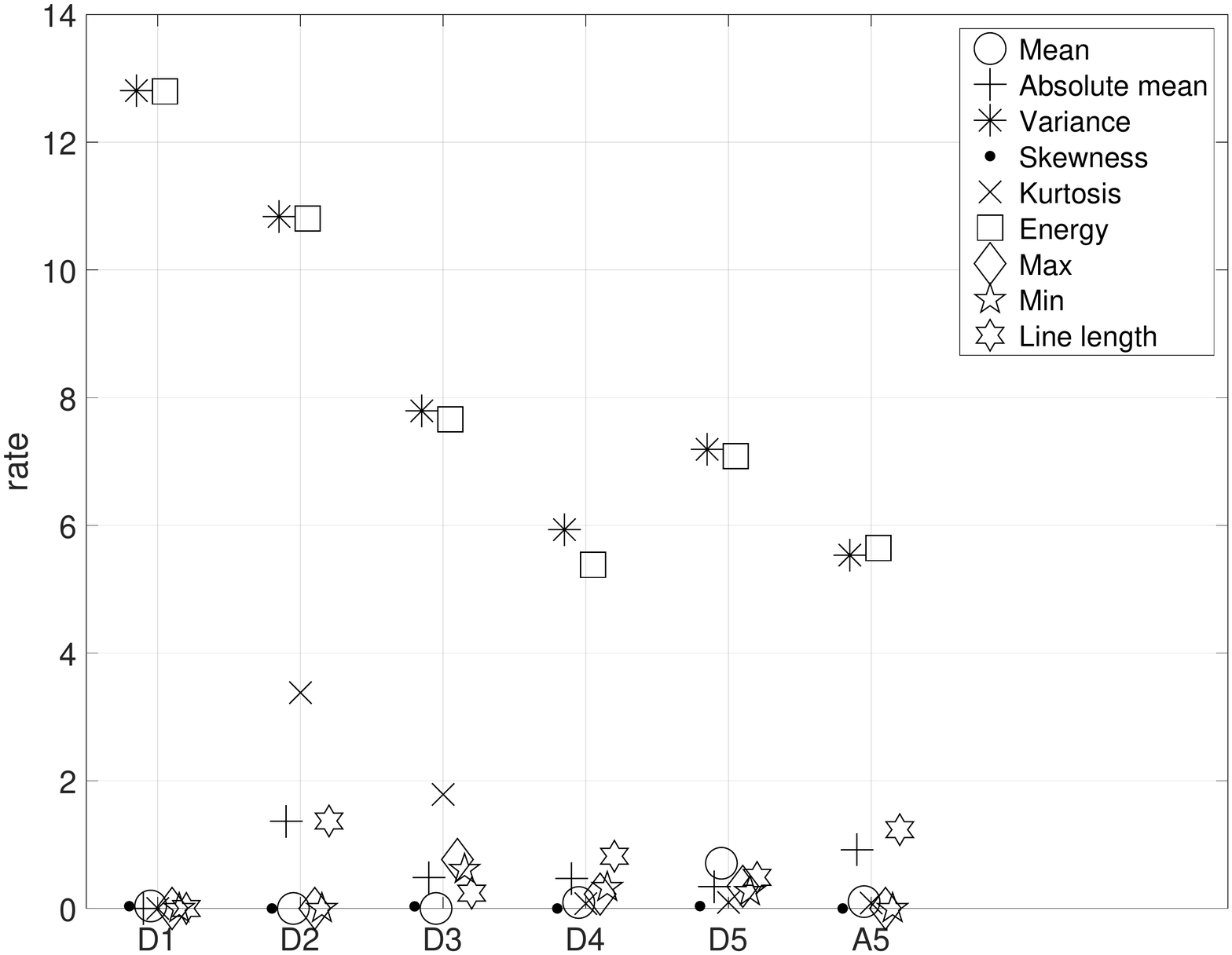}
    \caption{Improvement rate calculated from the \emph{right} hemisphere.}
    \label{subfig:imp_right_tf}
    \end{subfigure}
	\caption{Improvement rate based on the Bayesian method of time-frequency domain features calculated from \textbf{DWT} using Daubechies 4 wavelet.}
	\label{fig:imp_all}
\end{figure}
From the chosen records, there were 263,424 samples in the normal group and 4,677 epochs belonging to the seizure group, and so $\mathrm{err_{0}} = 4,677/(263,424+4,677) = 0.0174$. The features with an improvement rate higher than 4.5\% were considered as significant. \Cref{tab:imp_t_f} shows the Bayes error rate and improvement rate of each TDF and FDF. The results from this data set showed that most features achieved almost the same Bayes errors that were close to $\mathrm{err_0}$, except for the \emph{variance, energy}, \emph{NE} and \emph{ShEn} that obtained high improvement rates of 4.77\% to 10.07\%. \Cref{fig:imp_all} displays the improvement rates of all TFDFs in each wavelet decomposition level. Note that D1, D2, D3, D4, D5, and A5 represent sub-bands from which the features are extracted. Overall, the results showed that the \emph{variance} and \emph{energy} of the wavelet coefficients in all decomposition levels yielded relatively high improvement rates, compared to other features computed on the coefficients of other levels. Specifically, \emph{energy} from level D1 of the left half brain accomplished the highest improvement rate of 13.51\%. \emph{Line length} from levels D5 and A5, and \emph{kurtosis} from D1 of the left hemisphere also achieved significant improvement rates.
We found that the most significant features were related to amplitudes and variations of the signals, such as variance, energy, and NE.
Additionally, features that can capture changes in amplitude, frequency, and rhythmicity gain some improvement, since there is continuous evolution of amplitude, frequency, and rhythms during seizure activities compared to the background~\cite{pauri1992long}.
On the other hand, FDFs do not help improve the performance from the baseline because, in this data set, there are artifacts causing the probability of the seizure occurrence to be less than that of the other class.

\subsection{Feature redundancy analysis}
\label{sec:feat_redun}
\begin{figure}[h!]
	\centering
	\includegraphics[width=0.5\linewidth]{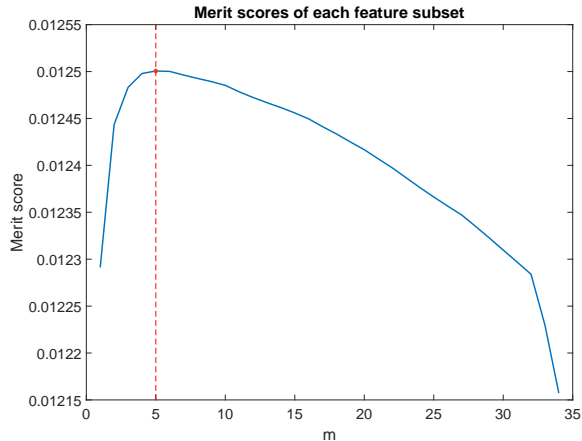}
	\caption{Merit scores of feature subsets. The subset size achieving the highest merit score is 5.}
	\label{fig:cfs_trend}
\end{figure}

The significant features achieving the improvement rate higher than 4.5\% (in total, 34 features) were then applied to the CFS algorithm to examine their redundancy. \Cref{fig:cfs_trend} shows that the merit score initially increases as $m$ (the number of features in the subset) increases until the feature subset contains five features, with a maximum score of $1.25\times 10^{-2}$. As $m$ increases beyond five, the merit score decreases. 

\begin{figure}[h!]
	\centering
	\includegraphics[width=0.5\linewidth]{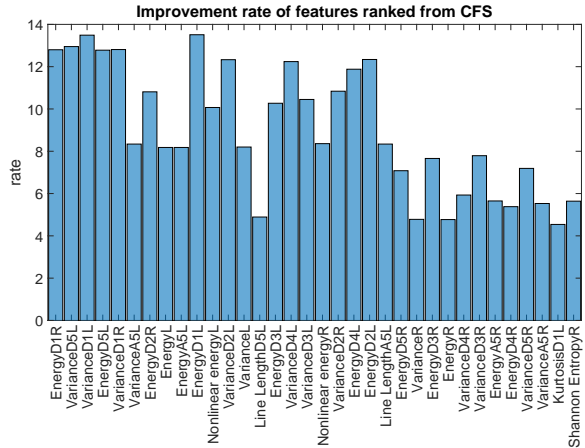}
	\caption{The features ranked by CFS and their improvement rates. All features in the optimal subset also obtained high improvement rates.}
	\label{fig:imp_cfs}
\end{figure}
In addition, the features and their improvement rates in the final feature subset were ranked in the descending order by the CFS as shown in \Cref{fig:imp_cfs}, where L and R stand for features computed from the left side and right side, respectively.
The features in the optimal subset were \emph{EnergyD1R, VarianceD5L, VarianceD1L, EnergyD5L}, and \emph{VarianceD1R}.
Accordingly, these five features also achieved relatively high improvement rates among the  significant features.
As shown in \Cref{fig:cfs_trend}, the merit scores of the optimal feature subset of each size were not that different.
This indicated that the results of feature significance from our experiments also agree with the outputs of the CFS.
The results in~\cite{aarabi2006automated} showed that the classification accuracy was not increased when other features were added to the optimal feature subset.
Moreover, these five features require just only $\mathcal{O}(N)$ for the total computation.
Hence, we suggest that these five features should be at least used as features in the AESD because of their high improvement rates and low computational complexity.

\section{Conclusions}
\label{sec:conc}
This paper aimed to review details of features in epileptic seizure detection using EEG signals.
We provided mathematical descriptions and computations of features in detail to reduce inconsistencies of some complicated feature definitions that have appeared in the literature.
We also summarized feature usages and intuitive meanings for detecting seizures according to their algorithms. 
Based on our review, deep learning techniques were used in combination with raw EEG signals or coefficients of a transformation, while shallow neural networks always required features as inputs. A single statistical parameter (as a feature) was never used but combinations of them were applied.
Such common combinations, including energy and entropy, were applied with a classification from ML approaches, while other types of features, such as entropy and fractal dimension, had sometimes been used independently. It was evident that the feature most widely used to quantify a relation with amplitude was energy.

Furthermore, we conducted two experiments to verify the significance of each widely used feature from the time, frequency, and time-frequency domains, and to analyze the redundancy of those features.
Based on our experiment using the standard Bayes classifier, we concluded that variance, energy, NE, and ShEn extracted from raw EEG signals were useful for distinguishing between seizure and normal EEG signals, and could improve the detection performance from the baseline condition.
However, other TDFs and FDFs gave insignificant outcomes for the detection of seizures.
Furthermore, features extracted from the DWT coefficients from some specific levels also provided intrinsically useful information about seizure that could not be achieved from raw EEG signals.
Moreover, based on redundancy analysis, the top significant features selected from CFS were the energy and variance from the DWT coefficients of some decomposition levels.
Therefore, the energy and variance calculated from the DWT coefficients were typically recommended to be used in both online and offline AESD due to their significance and low computational complexity.

\section*{Acknowledgement}
This work is financially supported by the 100th Anniversary Chulalongkorn University Fund for Doctoral Scholarship and the 90th Anniversary of Chulalongkorn University Fund (Ratchada-phiseksomphot Endowment Fund).

\bibliographystyle{alpha} 
\bibliography{eeg_references}

\newcommand{\etalchar}[1]{$^{#1}$}
\begin{thebibliography}{HWdVR08}

\bibitem[AESAA15]{alotaiby2015review}
T.~Alotaiby, F.E.A. El-Samie, S.A. Alshebeili, and I.~Ahmad.
\newblock A review of channel selection algorithms for {EEG} signal processing.
\newblock {\em EURASIP Journal on Advances in Signal Processing},
  2015(1):66--86, 2015.

\bibitem[AFS{\etalchar{+}}15]{acharya2015application}
U.R. Acharya, H.~Fujita, V.K. Sudarshan, S.~Bhat, and J.E.W. Koh.
\newblock Application of entropies for automated diagnosis of epilepsy using
  {EEG} signals: A review.
\newblock {\em Knowledge-Based Systems}, 88:85--96, 2015.

\bibitem[AHD{\etalchar{+}}18]{acharya2018characterization}
U.~R. Acharya, Y.~Hagiwara, S.~N. Deshpande, S.~Suren, J.~E.~W. Koh, S.~L. Oh,
  N.~Arunkumar, E.~J. Ciaccio, and C.~M. Lim.
\newblock Characterization of focal {EEG} signals: A review.
\newblock {\em Future Generation Computer Systems}, 91:290--299, 2018.

\bibitem[AJG17]{anand2017automatic}
S.~Anand, S.~Jaiswal, and P.K. Ghosh.
\newblock Automatic focal epileptic seizure detection in {EEG} signals.
\newblock In {\em Proceedings of the 2017 IEEE International WIE Conference on
  Electrical and Computer Engineering}, pages 103--107. IEEE, 2017.

\bibitem[AKS18]{alickovic2018performance}
E.~Alickovic, J.~Kevric, and A.~Subasi.
\newblock Performance evaluation of empirical mode decomposition, discrete
  wavelet transform, and wavelet packed decomposition for automated epileptic
  seizure detection and prediction.
\newblock {\em Biomedical Signal Processing and Control}, 39:94--102, 2018.

\bibitem[ALM{\etalchar{+}}01]{andrzejak2001indications}
R.G. Andrzejak, K.~Lehnertz, F.~Mormann, C.~Rieke, P.~David, and C.E. Elger.
\newblock Indications of nonlinear deterministic and finite-dimensional
  structures in time series of brain electrical activity: Dependence on
  recording region and brain state.
\newblock {\em Physical Review. E}, 64(6):061907, 2001.

\bibitem[ANRS17]{assi2017towards}
E.B. Assi, D.K. Nguyen, S.~Rihana, and M.~Sawan.
\newblock Towards accurate prediction of epileptic seizures: A review.
\newblock {\em Biomedical Signal Processing and Control}, 34:144--157, 2017.

\bibitem[AOH{\etalchar{+}}18]{acharya2018deep}
U.R. Acharya, S.L. Oh, Y.~Hagiwara, J.H. Tan, and H.~Adeli.
\newblock Deep convolutional neural network for the automated detection and
  diagnosis of seizure using {EEG} signals.
\newblock {\em Computers in Biology and Medicine}, 100(1):270--278, 2018.

\bibitem[AS16]{ammar2016seizure}
S.~Ammar and M.~Senouci.
\newblock Seizure detection with single-channel {EEG} using extreme learning
  machine.
\newblock In {\em Proceedings of the 17th International Conference on Sciences
  and Techniques of Automatic Control and Computer Engineering}, pages
  776--779. IEEE, 2016.

\bibitem[ASR12]{andrzejak2012nonrandomness}
R.~G. Andrzejak, K.~Schindler, and C.~Rummel.
\newblock Nonrandomness, nonlinear dependence, and nonstationarity of
  electroencephalographic recordings from epilepsy patients.
\newblock {\em Physical Review E}, 86(4):046206, 2012.

\bibitem[ASS{\etalchar{+}}13]{acharya2013automated}
U.R. Acharya, S.V. Sree, G.~Swapna, R.J. Martis, and J.S. Suri.
\newblock Automated {EEG} analysis of epilepsy: A review.
\newblock {\em Knowledge-Based Systems}, 45:147--165, 2013.

\bibitem[ASSK16]{ahmad2016seizure}
M.Z. Ahmad, M.~Saeed, S.~Saleem, and A.M. Kamboh.
\newblock Seizure detection using {EEG}: A survey of different techniques.
\newblock In {\em Proceedings of the 2016 International Conference on Emerging
  Technologies}, pages 1--6. IEEE, 2016.

\bibitem[ATE10]{altunay2010epileptic}
S.~Altunay, Z.~Telatar, and O.~Erogul.
\newblock Epileptic {EEG} detection using the linear prediction error energy.
\newblock {\em Expert Systems with Applications}, 37(8):5661--5665, 2010.

\bibitem[AWG06]{aarabi2006automated}
A.~Aarabi, F.~Wallois, and R.~Grebe.
\newblock Automated neonatal seizure detection: a multistage classification
  system through feature selection based on relevance and redundancy analysis.
\newblock {\em Clinical Neurophysiology}, 117(2):328--340, 2006.

\bibitem[BO18]{boashash2018designing}
B.~Boashash and S.~Ouelha.
\newblock Designing high-resolution time--frequency and time--scale
  distributions for the analysis and classification of non-stationary signals:
  a tutorial review with a comparison of features performance.
\newblock {\em Digital Signal Processing}, 77:120--152, 2018.

\bibitem[BP02]{bandt2002permutation}
C.~Bandt and B.~Pompe.
\newblock Permutation entropy: a natural complexity measure for time series.
\newblock {\em Physical Review Letters}, 88(17):174102, 2002.

\bibitem[BS12]{bryce2012revisiting}
R.M. Bryce and K.B. Sprague.
\newblock Revisiting detrended fluctuation analysis.
\newblock {\em Scientific Reports}, 2:315--320, 2012.

\bibitem[BTRD15]{bandarabadi2015epileptic}
M.~Bandarabadi, C.A. Teixeira, J.~Rasekhi, and A.~Dourado.
\newblock Epileptic seizure prediction using relative spectral power features.
\newblock {\em Clinical Neurophysiology}, 126(2):237--248, 2015.

\bibitem[CCS{\etalchar{+}}18]{chakraborti2018machine}
S.~Chakraborti, A.~Choudhary, A.~Singh, R.~Kumar, and A.~Swetapadma.
\newblock A machine learning based method to detect epilepsy.
\newblock {\em International Journal of Information Technology}, pages 1--7,
  2018.

\bibitem[Che14]{chen2014automatic}
G.~Chen.
\newblock Automatic {EEG} seizure detection using dual-tree complex
  wavelet-{Fourier} features.
\newblock {\em Expert Systems with Applications}, 41(5):2391--2394, 2014.

\bibitem[CLRS09]{cormen2009introduction}
T.~H. Cormen, Charles~E. Leiserson, R.~L. Rivest, and C.~Stein.
\newblock {\em Introduction to Algorithms}.
\newblock MIT press, 2009.

\bibitem[CWXY07]{chen2007characterization}
W.~Chen, Z.~Wang, H.~Xie, and W.~Yu.
\newblock Characterization of surface {EMG} signal based on fuzzy entropy.
\newblock {\em IEEE Transactions on Neural Systems and Rehabilitation
  Engineering}, 15(2):266--272, 2007.

\bibitem[DGL13]{devroye2013probabilistic}
L.~Devroye, L.~Gy{\"o}rfi, and G.~Lugosi.
\newblock {\em A Probabilistic Theory of Pattern Recognition}, volume~31.
\newblock Springer Science \& Business Media, 2013.

\bibitem[DJCF93]{dingle1993multistage}
A.A. Dingle, R.D. Jones, G.J. Carroll, and W.R. Fright.
\newblock A multistage system to detect epileptiform activity in the {EEG}.
\newblock {\em IEEE Transactions on Biomedical Engineering}, 40(12):1260--1268,
  1993.

\bibitem[EET{\etalchar{+}}01]{esteller2001line}
R.~Esteller, J.~Echauz, T.~Tcheng, B.~Litt, and B.~Pless.
\newblock Line length: an efficient feature for seizure onset detection.
\newblock In {\em Proceedings of the 23rd Annual International Conference of
  the IEEE Engineering in Medicine and Biology Society}, volume~2, pages
  1707--1710. IEEE, 2001.

\bibitem[FAA{\etalchar{+}}14]{fisher2014ilae}
R.S. Fisher, C.~Acevedo, A.~Arzimanoglou, A.~Bogacz, J.H. Cross, C.E. Elger,
  J.~Jr Engel, L.~Forsgren, J.A. French, M.~Glynn, et~al.
\newblock {ILAE} official report: a practical clinical definition of epilepsy.
\newblock {\em Epilepsia}, 55(4):475--482, 2014.

\bibitem[FAAA15]{faust2015wavelet}
O.~Faust, U.~R. Acharya, H.~Adeli, and A.~Adeli.
\newblock Wavelet-based {EEG} processing for computer-aided seizure detection
  and epilepsy diagnosis.
\newblock {\em Seizure}, 26:56--64, 2015.

\bibitem[Fal04]{falconer2004fractal}
K.~Falconer.
\newblock {\em Fractal Geometry: Mathematical Foundations and Applications}.
\newblock John Wiley \& Sons, 2004.

\bibitem[FAMS10]{faust2010automatic}
O.~Faust, U.R. Acharya, L.C. Min, and B.H.C. Sputh.
\newblock Automatic identification of epileptic and background {EEG} signals
  using frequency domain parameters.
\newblock {\em International Journal of Neural Systems}, 20(02):159--176, 2010.

\bibitem[FCKP13]{fadlallah2013weighted}
B.~Fadlallah, B.~Chen, A.~Keil, and J.~Pr{\'\i}ncipe.
\newblock Weighted-permutation entropy: A complexity measure for time series
  incorporating amplitude information.
\newblock {\em Physical Review. E}, 87(2):022911, 2013.

\bibitem[FHH{\etalchar{+}}16]{fergus2016machine}
P.~Fergus, A.~Hussain, D.~Hignett, D.~Al-Jumeily, K.~Abdel-Aziz, and H.~Hamdan.
\newblock A machine learning system for automated whole-brain seizure
  detection.
\newblock {\em Applied Computing and Informatics}, 12(1):70--89, 2016.

\bibitem[FHH{\etalchar{+}}18]{faust2018deep}
O.~Faust, Y.~Hagiwara, T.J. Hong, O.S. Lih, and U.R. Acharya.
\newblock Deep learning for healthcare applications based on physiological
  signals: A review.
\newblock {\em Computer Methods and Programs in Biomedicine}, 161:1--13, 2018.

\bibitem[Fuk90]{fukunaga2013introduction}
K.~Fukunaga.
\newblock {\em Introduction to Statistical Pattern Recognition}.
\newblock Academic Press, Inc., second edition, 1990.

\bibitem[F{\v{Z}}13]{fele2013faster}
G.~Fele-{\v{Z}}or{\v{z}}.
\newblock A faster algorithm for calculating the sample entropy.
\newblock In {\em Proceedings of the Middle-European Conference on Applied
  Theoretical Computer Science, Slovenia}, 2013.

\bibitem[GAG{\etalchar{+}}00]{Goldbergere215}
A.L. Goldberger, L.A.N. Amaral, L.~Glass, J.M. Hausdorff, P.C. Ivanov, R.G.
  Mark, J.E. Mietus, G.B. Moody, C.~Peng, and H.E. Stanley.
\newblock Physio{B}ank, {P}hysio{T}oolkit, and {P}hysio{N}et.
\newblock {\em Circulation}, 101(23):e215--e220, 2000.

\bibitem[GFM{\etalchar{+}}08]{greene2008comparison}
B.R. Greene, S.~Faul, W.P. Marnane, G.~Lightbody, I.~Korotchikova, and G.B.
  Boylan.
\newblock A comparison of quantitative {EEG} features for neonatal seizure
  detection.
\newblock {\em Clinical Neurophysiology}, 119(6):1248--1261, 2008.

\bibitem[GFZR97]{gotman1997automatic}
J.~Gotman, D.~Flanagan, J.~Zhang, and B.~Rosenblatt.
\newblock Automatic seizure detection in the newborn: methods and initial
  evaluation.
\newblock {\em Electroencephalography and Clinical Neurophysiology},
  103(3):356--362, 1997.

\bibitem[GG76]{gotman1976automatic}
J.~Gotman and P.~Gloor.
\newblock Automatic recognition and quantification of interictal epileptic
  activity in the human scalp {EEG}.
\newblock {\em Electroencephalography and Clinical Neurophysiology},
  41(5):513--529, 1976.

\bibitem[Got82]{gotman1982automatic}
J.~Gotman.
\newblock Automatic recognition of epileptic seizures in the {EEG}.
\newblock {\em Electroencephalography and Clinical Neurophysiology},
  54(5):530--540, 1982.

\bibitem[GRD{\etalchar{+}}10]{guo2010automatic}
L.~Guo, D.~Rivero, J.~Dorado, J.R. Rabunal, and A.~Pazos.
\newblock Automatic epileptic seizure detection in {EEGs} based on line length
  feature and artificial neural networks.
\newblock {\em Journal of Neuroscience Methods}, 191(1):101--109, 2010.

\bibitem[GRP10]{guo2010epileptic}
L.~Guo, D.~Rivero, and A.~Pazos.
\newblock Epileptic seizure detection using multiwavelet transform based
  approximate entropy and artificial neural networks.
\newblock {\em Journal of Neuroscience Methods}, 193(1):156--163, 2010.

\bibitem[GSK18]{gupta2018novel}
A.~Gupta, P.~Singh, and M.~Karlekar.
\newblock A novel signal modeling approach for classification of seizure and
  seizure-free {EEG} signals.
\newblock {\em IEEE Transactions on Neural Systems and Rehabilitation
  Engineering}, 26(5):925--935, 2018.

\bibitem[Hal99]{hall1999correlation}
M.A. Hall.
\newblock {\em Correlation-based feature selection for machine learning}.
\newblock PhD thesis, University of Waikato Hamilton, New Zealand, 1999.

\bibitem[Hig88]{higuchi1988approach}
T.~Higuchi.
\newblock Approach to an irregular time series on the basis of the fractal
  theory.
\newblock {\em Physica D: Nonlinear Phenomena}, 31(2):277--283, 1988.

\bibitem[Hjo70]{hjorth1970eeg}
B.~Hjorth.
\newblock {EEG} analysis based on time domain properties.
\newblock {\em Electroencephalography and Clinical Neurophysiology},
  29(3):306--310, 1970.

\bibitem[HS97]{hall1997feature}
M.A. Hall and L.A. Smith.
\newblock Feature subset selection: a correlation based filter approach.
\newblock In {\em Proceedings of the 4th International Conference on Neural
  Information Processing and Intelligent Information Systems}, pages 855--858.
  Springer, 1997.

\bibitem[Hua14]{huang2014hilbert}
N.E. Huang.
\newblock {\em Hilbert-{Huang} transform and its applications}, volume~16.
\newblock World Scientific, 2014.

\bibitem[Hur51]{hurst1951long}
H.E. Hurst.
\newblock Long-term storage capacity of reservoirs.
\newblock {\em Transactions of the American Society of Civil Engineers},
  116:770--799, 1951.

\bibitem[HWdVR08]{hellstrom2008atlas}
L.~Hellstr{\"o}m-Westas, L.S. de~Vries, and I.~Ros{\'e}n.
\newblock {\em An atlas of amplitude-integrated {EEGs} in the newborn}.
\newblock CRC Press, 2008.

\bibitem[HZS06]{huang2006extreme}
G.~Huang, Q.~Zhu, and C.~Siew.
\newblock Extreme learning machine: theory and applications.
\newblock {\em Neurocomputing}, 70(1-3):489--501, 2006.

\bibitem[Jan17a]{janjarasjitt2017epileptic}
S.~Janjarasjitt.
\newblock Epileptic seizure classifications of single-channel scalp {EEG} data
  using wavelet-based features and svm.
\newblock {\em Medical \& Biological Engineering \& Computing},
  55(10):1743--1761, 2017.

\bibitem[Jan17b]{janjarasjitt2017performance}
S.~Janjarasjitt.
\newblock Performance of epileptic single-channel scalp {EEG} classifications
  using single wavelet-based features.
\newblock {\em Australasian Physical \& Engineering Sciences in Medicine},
  40(1):57--67, 2017.

\bibitem[JB17]{jaiswal2017local}
A.K. Jaiswal and H.~Banka.
\newblock Local pattern transformation based feature extraction techniques for
  classification of epileptic {EEG} signals.
\newblock {\em Biomedical Signal Processing and Control}, 34:81--92, 2017.

\bibitem[Kai90]{kaiser1990simple}
J.F. Kaiser.
\newblock On a simple algorithm to calculate the energy of a signal.
\newblock In {\em Proceedings of the International Conference on Acoustics,
  Speech, and Signal Processing}, pages 381--384. IEEE, 1990.

\bibitem[KKP15]{kumar2015classification}
T.S. Kumar, V.~Kanhangad, and R.B. Pachori.
\newblock Classification of seizure and seizure-free {EEG} signals using local
  binary patterns.
\newblock {\em Biomedical Signal Processing and Control}, 15:33--40, 2015.

\bibitem[K{\v{S}}R{\v{S}}97]{kononenko1997overcoming}
I.~Kononenko, E.~{\v{S}}imec, and M.~Robnik-{\v{S}}ikonja.
\newblock Overcoming the myopia of inductive learning algorithms with
  {RELIEFF}.
\newblock {\em Applied Intelligence}, 7(1):39--55, 1997.

\bibitem[LCZ17a]{li2017automatic}
M.~Li, W.~Chen, and T.~Zhang.
\newblock Automatic epileptic {EEG} detection using {DT-CWT}-based non-linear
  features.
\newblock {\em Biomedical Signal Processing and Control}, 34:114--125, 2017.

\bibitem[LCZ17b]{li2017classification}
M.~Li, W.~Chen, and T.~Zhang.
\newblock Classification of epilepsy {EEG} signals using {DWT}-based envelope
  analysis and neural network ensemble.
\newblock {\em Biomedical Signal Processing and Control}, 31:357--365, 2017.

\bibitem[LKY{\etalchar{+}}18]{li2018detection}
P.~Li, C.~Karmakar, J.~Yearwood, S.~Venkatesh, M.~Palaniswami, and C.~Liu.
\newblock Detection of epileptic seizure based on entropy analysis of
  short-term {EEG}.
\newblock {\em PloS one}, 13(3):e0193691, 2018.

\bibitem[LLL{\etalchar{+}}15]{li2015assessing}
P.~Li, C.~Liu, K.~Li, D.~Zheng, C.~Liu, and Y.~Hou.
\newblock Assessing the complexity of short-term heartbeat interval series by
  distribution entropy.
\newblock {\em Medical {\&} Biological Engineering {\&} Computing},
  53(1):77--87, Jan 2015.

\bibitem[LYLO14]{li2014using}
J.~Li, J.~Yan, X.~Liu, and G.~Ouyang.
\newblock Using permutation entropy to measure the changes in {EEG} signals
  during absence seizures.
\newblock {\em Entropy}, 16(6):3049--3061, 2014.

\bibitem[Man08]{manis2008fast}
G.~Manis.
\newblock Fast computation of approximate entropy.
\newblock {\em Computer Methods and Programs in Biomedicine}, 91(1):48--54,
  2008.

\bibitem[MAS18]{manis2018low}
G.~Manis, M.~Aktaruzzaman, and R.~Sassi.
\newblock Low computational cost for sample entropy.
\newblock {\em Entropy}, 20(1):61--75, 2018.

\bibitem[MPS69]{maynard1969device}
D.E. Maynard, P.F. Prior, and D.F. Scott.
\newblock Device for continuous monitoring of cerebral activity in resuscitated
  patients.
\newblock {\em British Medical Journal}, 4(5682):545--546, 1969.

\bibitem[MZCC17]{mursalin2017automated}
M.~Mursalin, Y.~Zhang, Y.~Chen, and N.V. Chawla.
\newblock Automated epileptic seizure detection using improved
  correlation-based feature selection with random forest classifier.
\newblock {\em Neurocomputing}, 241:204--214, 2017.

\bibitem[Oca09]{ocak2009automatic}
H.~Ocak.
\newblock Automatic detection of epileptic seizures in {EEG} using discrete
  wavelet transform and approximate entropy.
\newblock {\em Expert Systems with Applications}, 36(2):2027--2036, 2009.

\bibitem[OLC{\etalchar{+}}09]{orosco2009epileptic}
L.~Orosco, E.~Laciar, A.G Correa, A.~Torres, and J.P. Graffigna.
\newblock An epileptic seizures detection algorithm based on the empirical mode
  decomposition of {EEG}.
\newblock In {\em Proceedings of the Annual International Conference of the
  IEEE Engineering in Medicine and Biology Society}, pages 2651--2654. IEEE,
  2009.

\bibitem[Par62]{parzen1962estimation}
E.~Parzen.
\newblock On estimation of a probability density function and mode.
\newblock {\em The Annals of Mathematical Statistics}, 33(3):1065--1076, 1962.

\bibitem[per]{persyst2019seizure}
Seizure {D}etection.
\newblock \url{https://www.persyst.com/technology/seizure-detection/}.
\newblock Accessed: 2019-4-25.

\bibitem[PG07]{polat2007classification}
K.~Polat and S.~G{\"u}ne{\c{s}}.
\newblock Classification of epileptiform {EEG} using a hybrid system based on
  decision tree classifier and fast {Fourier} transform.
\newblock {\em Applied Mathematics and Computation}, 187(2):1017--1026, 2007.

\bibitem[Pin91]{pincus1991approximate}
S.M. Pincus.
\newblock Approximate entropy as a measure of system complexity.
\newblock {\em Proceedings of the National Academy of Sciences},
  88(6):2297--2301, 1991.

\bibitem[PPCE92]{pauri1992long}
F.~Pauri, F.~Pierelli, G.~Chatrian, and W.W. Erdly.
\newblock Long-term {EEG}-video-audio monitoring: computer detection of focal
  {EEG} seizure patterns.
\newblock {\em Electroencephalography and Clinical Neurophysiology},
  82(1):1--9, 1992.

\bibitem[RM00]{richman2000physiological}
J.S. Richman and J.R. Moorman.
\newblock Physiological time-series analysis using approximate entropy and
  sample entropy.
\newblock {\em American Journal of Physiology-Heart and Circulatory
  Physiology}, 278(6):H2039--H2049, 2000.

\bibitem[RPR99]{roberts1999temporal}
S.J. Roberts, W.~Penny, and I.~Rezek.
\newblock Temporal and spatial complexity measures for electroencephalogram
  based brain-computer interfacing.
\newblock {\em Medical \& Biological Engineering \& Computing}, 37(1):93--98,
  1999.

\bibitem[SEC{\etalchar{+}}04]{shoeb2004patient}
A.H. Shoeb, H.~Edwards, J.~Connolly, B.~Bourgeois, S.T. Treves, and J.~Guttag.
\newblock Patient-specific seizure onset detection.
\newblock {\em Epilepsy \& Behavior}, 5(4):483--498, 2004.

\bibitem[SG10a]{shoeb2010application}
A.H. Shoeb and J.~Guttag.
\newblock Application of machine learning to epileptic seizure detection.
\newblock In {\em Proceedings of the 27th International Conference on Machine
  Learning}, pages 975--982, 2010.

\bibitem[SG10b]{subasi2010eeg}
A.~Subasi and M.I. Gursoy.
\newblock {EEG} signal classification using {PCA}, {ICA}, {LDA} and support
  vector machines.
\newblock {\em Expert Systems with Applications}, 37(12):8659--8666, 2010.

\bibitem[Sha48]{shannon1948mathematical}
C.E. Shannon.
\newblock A mathematical theory of communication.
\newblock {\em Bell System Technical Journal}, 27(3):379--423, 1948.

\bibitem[Sho09]{shoeb2009application}
A.H. Shoeb.
\newblock {\em Application of machine learning to epileptic seizure onset
  detection and treatment}.
\newblock PhD thesis, Massachusetts Institute of Technology, 2009.

\bibitem[SKG15]{samiee2015epileptic}
K.~Samiee, P.~Kovacs, and M.~Gabbouj.
\newblock Epileptic seizure classification of {EEG} time-series using rational
  discrete short-time {Fourier} transform.
\newblock {\em IEEE Transactions on Biomedical Engineering}, 62(2):541--552,
  2015.

\bibitem[SLUC15]{satirasethawong2015amplitude}
C.~Satirasethawong, A.~Lek-Uthai, and K.~Chomtho.
\newblock Amplitude-integrated {EEG} processing and its performance for
  automatic seizure detection.
\newblock In {\em Proceedings of the 2015 IEEE International Conference on
  Signal and Image Processing Applications}, pages 551--556. IEEE, 2015.

\bibitem[SNM14]{salem2014epileptic}
O.~Salem, A.~Naseem, and A.~Mehaoua.
\newblock Epileptic seizure detection from {EEG} signal using discrete wavelet
  transform and ant colony classifier.
\newblock In {\em Proceedings of the 2014 IEEE International Conference on
  Communications}, pages 3529--3534. IEEE, 2014.

\bibitem[TBB{\etalchar{+}}11]{thurman2011standards}
D.J. Thurman, E.~Beghi, C.E. Begley, A.T. Berg, J.R. Buchhalter, D.~Ding, D.C.
  Hesdorffer, W.A. Hauser, L.~Kazis, R.~Kobau, et~al.
\newblock Standards for epidemiologic studies and surveillance of epilepsy.
\newblock {\em Epilepsia}, 52(7):2--26, 2011.

\bibitem[TTF07]{tzallas2007automatic}
A.T. Tzallas, M.G. Tsipouras, and D.I. Fotiadis.
\newblock Automatic seizure detection based on time-frequency analysis and
  artificial neural networks.
\newblock {\em Computational Intelligence and Neuroscience}, 2007, 2007.

\bibitem[TTF09]{tzallas2009epileptic}
A.T. Tzallas, M.G. Tsipouras, and D.I. Fotiadis.
\newblock Epileptic seizure detection in {EEGs} using time--frequency analysis.
\newblock {\em IEEE Transactions on Information Technology in Biomedicine},
  13(5):703--710, 2009.

\bibitem[TTM{\etalchar{+}}11a]{temko2011eeg}
A.~Temko, E.~Thomas, W.~Marnane, G.~Lightbody, and G.~Boylan.
\newblock {EEG}-based neonatal seizure detection with support vector machines.
\newblock {\em Clinical Neurophysiology}, 122(3):464--473, 2011.

\bibitem[TTM{\etalchar{+}}11b]{temko2011performance}
A.~Temko, E.~Thomas, W.~Marnane, G.~Lightbody, and G.B. Boylan.
\newblock Performance assessment for {EEG}-based neonatal seizure detectors.
\newblock {\em Clinical Neurophysiology}, 122(3):474--482, 2011.

\bibitem[TYK16]{tawfik2016hybrid}
N.S. Tawfik, S.M. Youssef, and M.~Kholief.
\newblock A hybrid automated detection of epileptic seizures in {EEG} records.
\newblock {\em Computers \& Electrical Engineering}, 53:177--190, 2016.

\bibitem[VI17]{vidyaratne2017real}
L.S. Vidyaratne and K.M. Iftekharuddin.
\newblock Real-time epileptic seizure detection using {EEG}.
\newblock {\em IEEE Transactions on Neural Systems and Rehabilitation
  Engineering}, 25(11):2146--2156, 2017.

\bibitem[WMX11]{wang2011best}
D.~Wang, D.~Miao, and C.~Xie.
\newblock Best basis-based wavelet packet entropy feature extraction and
  hierarchical {EEG} classification for epileptic detection.
\newblock {\em Expert Systems with Applications}, 38(11):14314--14320, 2011.

\bibitem[YL04]{yu2004efficient}
L.~Yu and H.~Liu.
\newblock Efficient feature selection via analysis of relevance and redundancy.
\newblock {\em Journal of Machine Learning Research}, 5(Oct):1205--1224, 2004.

\bibitem[YXJZ17]{yuan2017multi}
Y.~Yuan, G.~Xun, K.~Jia, and A.~Zhang.
\newblock A multi-view deep learning method for epileptic seizure detection
  using short-time fourier transform.
\newblock In {\em Proceedings of the 8th ACM International Conference on
  Bioinformatics, Computational Biology, and Health Informatics}, pages
  213--222. ACM, 2017.

\bibitem[ZMS{\etalchar{+}}10]{zhao2010advancing}
Z.~Zhao, F.~Morstatter, S.~Sharma, S.~Alelyani, A.~Anand, and H.~Liu.
\newblock Advancing feature selection research.
\newblock {\em ASU Feature Selection Repository}, pages 1--28, 2010.

\end{thebibliography}

\end{document}